\documentclass[twocolumn]{article}
\let\ifapj\iffalse
\let\ifarxiv\iftrue
\let\iflocal\iffalse
\usepackage{ifxetex,ifluatex}
\ifxetex\else\ifluatex\else\usepackage[utf8]{inputenc}\fi\fi
\ifarxiv\usepackage{font-termes}\fi
\iflocal\usepackage{font-termes}\fi
\makeatletter
\@ifpackageloaded{unicode-math}{\let\abbrev\symsf}{\let\abbrev\mathsf}
\makeatother
\usepackage{etoolbox,ifthen}
\ifapj
  \usepackage[strict]{patch-apj}
  \usepackage{acro,appendix,booktabs,csquotes,hyperref,siunitx}
  \hypersetup{hidelinks}
\fi
\ifboolexpr{bool{arxiv} or bool{local}}{
  \usepackage{booktabs,siunitx,tikz}
  \usepackage{basic,physics,astro}
  \usepackage[graytext]{draft}
  \usepackage[mode=biblatex, hyperref]{apjbib}
  \ExecuteBibliographyOptions{embedlinks}
  \addbibresource{mri.bib}
}{}
\ifluatex\hypersetup{pdfencoding=auto}\fi
\makeatletter
\ifarxiv\@ifpackageloaded{astro}{\let\ast@symit\textit}{}\fi
\makeatother
\DeclareAcronym{MHD} {short=MHD,  long=magnetohydrodynamic,           short-indefinite=an, short-plural=, long-plural=s}
\DeclareAcronym{MRI} {short=MRI,  long=magnetorotational instability, short-indefinite=an}
\DeclareAcronym{SMBH}{short=SMBH, long=supermassive black hole,       short-indefinite=an}
\DeclareAcronym{TDE} {short=TDE,  long=tidal disruption event}
\DeclareAcronym{GR}  {short=GR,   long=general relativistic}
\newcommand*\strcyl{\smash{T^{\hat\varphi}_{\hat R}}}
\newcommand*\strorb{\smash{T^{\hat\varphi}_{\hat\lambda}}}
\expandafter\def\csname editcolor1\endcsname{blue!60!cyan}
\ifapj\let\edit\relax\fi
\newcommand*\edit[2]{\textcolor{\csname editcolor#1\endcsname}{#2}}
\begin{document}

\title{Magnetorotational instability in eccentric disks}

\ifapj
  \author[0000-0001-5949-6109]{Chi-Ho Chan}
  \affiliation{Racah Institute of Physics, Hebrew University of Jerusalem,
  Jerusalem 91904, Israel}
  \affiliation{School of Physics and Astronomy, Tel Aviv University,
  Tel Aviv 69978, Israel}
  \author[0000-0002-2995-7717]{Julian~H. Krolik}
  \affiliation{Department of Physics and Astronomy, Johns Hopkins University,
  Baltimore, MD 21218, USA}
  \author[0000-0002-7964-5420]{Tsvi Piran}
  \affiliation{Racah Institute of Physics, Hebrew University of Jerusalem,
  Jerusalem 91904, Israel}
\fi

\ifboolexpr{bool{arxiv} or bool{local}}{
  \author[1,2]{Chi-Ho Chan}
  \author[3]{Julian~H. Krolik}
  \author[1]{Tsvi Piran}
  \affil[1]{Racah Institute of Physics, Hebrew University of Jerusalem,
  Jerusalem 91904, Israel}
  \affil[2]{School of Physics and Astronomy, Tel Aviv University,
  Tel Aviv 69978, Israel}
  \affil[3]{Department of Physics and Astronomy, Johns Hopkins University,
  Baltimore, MD 21218, USA}
}{}

\date{January 15, 2018}

\shorttitle{Eccentric magnetorotational instability}
\shortauthors{Chan et al.}
\pdftitle{Magnetorotational instability in eccentric disks}
\pdfauthors{Chi-Ho Chan, Julian H. Krolik, Tsvi Piran}

\begin{abstract}
Eccentric disks arise in such astrophysical contexts as \aclp*{TDE}, but it is
unknown whether the \ac{MRI}, which powers accretion in circular disks,
operates in eccentric disks as well. We examine the linear evolution of
unstratified, incompressible \ac{MRI} in an eccentric disk orbiting a point
mass. We consider vertical modes of wavenumber $k$ on a background flow with
uniform eccentricity $e$ and vertical Alfv\'en speed $v_\su A$ along an orbit
with mean motion $n$. We find two mode families, one with dominant magnetic
components, the other with dominant velocity components; the former is unstable
at $(1-e)^3f^2\lesssim3$, where $f\eqdef kv_\su A/n$, the latter at
$e\gtrsim0.8$. For $f^2\lesssim3$, \ac{MRI} behaves much like in circular
disks, but the growth per orbit declines slowly with increasing $e$; for
$f^2\gtrsim3$, modes grow by parametric amplification, which is resonant for
$0<e\ll1$. \Ac{MRI} growth and the attendant angular momentum and energy
transport happen chiefly near pericenter, where orbital shear dominates
magnetic tension.
\end{abstract}
\acresetall

\section{Introduction}

The \ac{MRI} is a powerful instability in weakly magnetized, differentially
rotating circular disks \citep{1991ApJ...376..214B, 1991ApJ...376..223H}. The
instability is most easily visualized by considering a disk threaded by a
vertical magnetic field with a small radial kink. Orbital shear pulls the kink
out toroidally, creating a correlation between the horizontal components of the
magnetic field perturbation, and also between those of the velocity
perturbation. The resulting Reynolds and Maxwell stresses transport angular
momentum outward; gas at smaller radii therefore moves inward, while gas at
larger radii moves outward. This stretches the initial kink further and allows
the instability to grow exponentially. The fact that \ac{MRI} grows as fast as
the orbital timescale guarantees its role as the mechanism by which ionized
disks accrete.

Disks can nevertheless be eccentric. Secular gravitational interaction in
eccentric binaries bestows forced eccentricity upon circumbinary and
circumobject disks \citep[e.g.,][]{2000ssd..book.....M}. Tidal forces in
circular binaries couple to circumobject disks through the $3\mathbin:1$ mean
motion resonance and allow small but finite free eccentricity to grow
exponentially \citep{1991ApJ...381..259L}. Viscous overstability
\citep{1978MNRAS.185..629K} amplifies small-scale eccentric perturbations in
isolated disks \citep[e.g.,][]{1994MNRAS.266..583L, 2001MNRAS.325..231O}.
Lastly, stars passing too close to \acp{SMBH}
\citep[e.g.,][]{1988Natur.333..523R} or planets grazing their host stars can be
tidally disrupted, and the bound debris can form an eccentric disk directly
\citep[e.g.,][]{2014ApJ...783...23G, 2015ApJ...804...85S, 2016MNRAS.455.2253B,
2016MNRAS.461.3760H}.

Shocks transfer angular momentum within the bound debris of \acp{TDE} around
\acp{SMBH}, particularly during the early stages of the event
\citep{1989ApJ...346L..13E, 1994ApJ...422..508K, 2013ApJ...767...25G,
2015ApJ...804...85S}. But since the condition of ideal \acp{MHD} requires
little ionization \citep{1994ApJ...421..163B, 1996ApJ...457..355G}, \ac{MRI} is
also likely active; angular momentum transport by \ac{MHD} stresses may then
control how the debris evolves. \Citet{2017MNRAS.467.1426S} showed that
near-apocenter parts of the orbit dominate angular momentum transport, while
near-pericenter parts dominate energy dissipation. They also argued that over
an orbit, the debris preferentially loses angular momentum rather than energy,
so it quickly plunges into the \ac{SMBH} without radiating much, in agreement
with observations. However, the effectiveness of angular momentum transport by
\ac{MHD} stresses depends on how fast \ac{MRI} grows, and no one has yet
considered how \ac{MRI} growth in eccentric disks might be different from
circular disks.

This article describes our first step toward understanding how \ac{MRI} behaves
in an eccentric disk orbiting a point mass. We study the linear evolution of
unstratified and incompressible (Boussinesq) \ac{MRI}; we call this
\definition{eccentric \ac{MRI}}, in contrast to \definition{circular \ac{MRI}},
its counterpart in circular disks. Both kinds of \ac{MRI} feed off orbital
shear; because orbital shear is radial and time-independent in circular disks
but oblique and time-varying in eccentric disks, we expect eccentric \ac{MRI}
to differ in nature from circular \ac{MRI}. It is not apparent whether
eccentric \ac{MRI} grows exponentially like circular \ac{MRI}, and how the
growth rates of circular and eccentric \ac{MRI} compare. More interestingly,
variation of orbital conditions along the orbit can destabilize inertial and
gravity modes in thin hydrodynamic disks through parametric resonance
\citep{2005A&A...432..743P}; a similar mechanism may destabilize their
magnetized counterparts in \ac{MHD} disks.

We present the linearized equations of eccentric \ac{MRI} and our method for
solving them in \cref{sec:methods}. We map out the growth per orbit of
eccentric \ac{MRI} as a function of eccentricity and perturbation wavenumber in
\cref{sec:growth}, describe qualitatively the time-evolution of unstable modes
in \cref{sec:phase}, and compute the angular momentum and energy fluxes due to
these modes in \cref{sec:stress}. We interpret our results with a toy model in
\cref{sec:model} and discuss their astrophysical importance in
\cref{sec:discussion}.

\section{Methods}
\label{sec:methods}

\subsection{Orbital and shearing-box coordinate systems}

\begin{figure}
\includegraphics{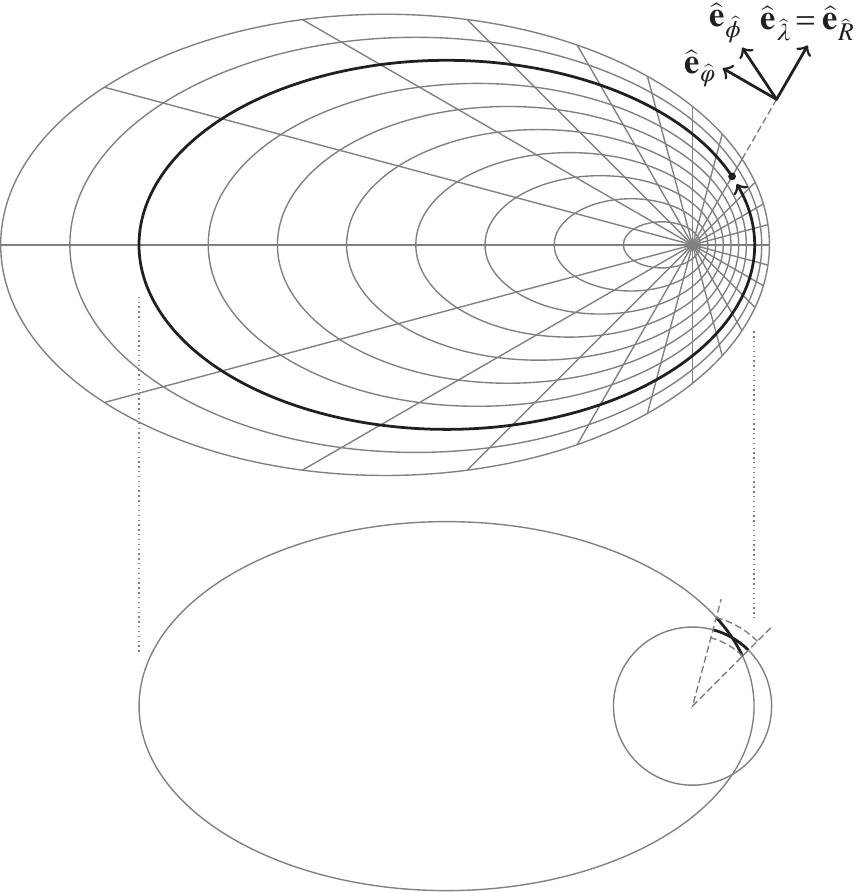}
\caption{\textit{Top half:} Coordinate curves of one realization of the orbital
coordinate system where constant\protect\nobreakdash-$\lambda$ contours have
the same orientation and eccentricity $e=0.8$. The thick contour is the orbit
of a reference particle; the perturbation at the reference particle evolves
according to \cref{eq:full MRI,eq:shearing-box MRI,eq:cylindrical
MRI,eq:reduced MRI} as the reference particle orbits a point mass at the
origin. The normalized shearing-box and cylindrical coordinate bases,
respectively $(\uvec e_{\hat\lambda},\uvec e_{\hat\phi})$ and $(\uvec e_{\hat
R},\uvec e_{\hat\varphi})$, at the present azimuth of the reference particle
are displayed in the corner. \textit{Bottom half:} Line segments used to
demonstrate why $\strcyl=\strorb$ (\cref{sec:stress}).}
\label{fig:coordinates}
\end{figure}

Our analysis is based on the framework laid out by
\ifapj\citet{2001MNRAS.325..231O} and \citet{2014MNRAS.445.2621O}\else
\citet{2001MNRAS.325..231O, 2014MNRAS.445.2621O}\fi.
\Citet{2001MNRAS.325..231O} introduced the \definition{orbital coordinate
system} $(\lambda,\phi)$, illustrated in the top half of
\cref{fig:coordinates}. A constant\nobreakdash-$\lambda$ contour is an ellipse
with semilatus rectum $\lambda$ and one focus at the origin, and $\phi$ is the
azimuth; the ellipses must vary slowly in orientation and eccentricity over
$\lambda$ so they do not intersect \citep{2014MNRAS.445.2621O}. The coordinate
system can be extended by adding a vertical coordinate $z$ perpendicular to the
plane of ellipses. Using standard methods of Riemannian differential geometry,
\citet{2014MNRAS.445.2621O} wrote down the components of the ideal \acp{MHD}
equations for adiabatic gas in this non-orthogonal coordinate system.

Because particles in the midplane orbit a point mass at the origin along
ellipses defining the orbital coordinate system, the orbital coordinate system
provides a foundation for extending the shearing box to eccentric disks. To do
so, \citet{2014MNRAS.445.2621O} chose some reference particle
$(\lambda_0,\theta(t))$ and defined a non-orthogonal, shearing-box coordinate
system $(\xi,\eta,\zeta)\eqdef(\lambda-\lambda_0,\phi-\theta(t),z)$ such that
$\xi/\lambda_0$, $\eta$, and $\zeta/\lambda_0$ are $\mathrelp\sim\epsilon$,
where $\epsilon\ll1$ is the disk aspect ratio; the reference particle and the
shearing-box coordinate basis, as well as another basis to be defined in
\cref{sec:cylindrical}, are shown in \cref{fig:coordinates}.
\Citet{2014MNRAS.445.2621O} obtained the velocity perturbation in the shearing
box by subtracting from the gas velocity in the inertial frame the velocity of
particles following coordinate ellipses, assuming that the velocity
perturbation is $\mathrelp\sim\epsilon$ times the particle velocity. Finally,
they subtracted from the time-derivative the contribution due to orbital
motion. This procedure gave them their Equations~(83), (84), (86),
and~(C4)--(C10), which are the \acp{MHD} equations nonlinear in the velocity
perturbation, and from which we derive \cref{eq:full MRI} below. We are
interested in how the perturbation at the same $(\lambda,\phi)$ as the
reference particle evolves, thus we can drop the subscript from $\lambda_0$
without ambiguity.

\subsection{Linearized \texorpdfstring{\acsp*{MHD}}{MHD} equations in
shearing-box coordinate basis}
\label{sec:MRI}

Henceforth we adopt the orbital coordinate system defined by aligned ellipses
of constant eccentricity $e$. We can convince ourselves that the velocity field
defined by particles orbiting the point mass along coordinate ellipses is
divergence-free by writing it out explicitly, but we can also see intuitively
why this is so: At pericenter, orbits are closer together but particles move
faster.

Because a divergence-free velocity field is incompressible, one solution of the
\acp{MHD} equations in the midplane is that density is a function of $\lambda$
only, pressure is uniform, magnetic field is vertical and uniform, and gas
travels along coordinate ellipses. We choose this solution as the unperturbed
background. We consider the particular case where the background density is
uniform and, as applies near the midplane, where vertical gravity can be
ignored; the latter condition means we are looking at the perturbation at fixed
$\zeta$ above the reference particle.

We specialize the \acp{MHD} equations of \citet{2014MNRAS.445.2621O} according
to these assumptions. Because we are considering the perturbation at the same
$(\lambda,\phi)$ as the reference particle, we elide terms proportional to
$\xi$ and $\eta$ in their Equation~(84). Because our background is uniform and
we ignore vertical stratification, we discard background spatial gradients and
vertical gravity from all their equations. Because the background flow is
divergence-free, we set its divergence $\Delta$ to zero in their
Equations~(83), (86), and~(C7)--(C9). We retain only terms that are first order
in perturbed quantities, and we replace
$(\partial_\xi,\partial_\eta,\partial_\zeta)$ by $(ik_\xi,\lambda
ik_\eta,ik_\zeta)$.

To arrive at our form of the linearized \acp{MHD} equations, we choose a
magnetic field unit that absorbs a factor of $(4\pi)^{-1/2}$. We let $\rho$,
$c_\su s$, and $B$ be the background density, adiabatic sound speed, and
vertical magnetic field respectively, and $v^\mu$ be the contravariant
components of the velocity perturbation in the shearing-box coordinate basis.
We also denote by $u$ the perturbed logarithmic density times $c_\su s$, and by
$w^\mu$ the magnetic field perturbation divided by $\rho^{1/2}$. The linearized
\acp{MHD} equations are then
\begin{equation}\label{eq:full MRI}
\ifapj\else\arraycolsep.54\arraycolsep\fi
\od{}M\begin{pmatrix} u \\ v^\xi \\ \lambda v^\eta \\ v^\zeta \\
  w^\xi \\ \lambda w^\eta \\ w^\zeta \end{pmatrix}=
\begin{pmatrix}
0 & -if^\su s_\xi & -if^\su s_\eta & -if^\su s_\zeta & 0 & 0 & 0 \\
\abbrev F^\su s_\lambda & 0 & \abbrev A & 0 &
  if^\su m_\zeta & 0 & \abbrev F^\su m_\lambda \\
\abbrev F^\su s_\phi & \abbrev B & \abbrev C & 0 &
  0 & if^\su m_\zeta & \abbrev F^\su m_\phi \\
-if^\su s_\zeta & 0 & 0 & 0 & 0 & 0 & 0 \\
0 & if^\su m_\zeta & 0 & 0 & 0 & 0 & 0 \\
0 & 0 & if^\su m_\zeta & 0 & \abbrev D & \abbrev E & 0 \\
0 & -if^\su m_\xi & -if^\su m_\eta & 0 & 0 & 0 & 0
\end{pmatrix}
\begin{pmatrix} u \\ v^\xi \\ \lambda v^\eta \\ v^\zeta \\
  w^\xi \\ \lambda w^\eta \\ w^\zeta \end{pmatrix},
\end{equation}
and the solenoidal condition for the magnetic field reads
\begin{equation}\label{eq:solenoidal}
if^\su m_\xi w^\xi+if^\su m_\eta(\lambda w^\eta)+if^\su m_\zeta w^\zeta=0.
\end{equation}
The time variable in \cref{eq:full MRI} is the mean anomaly $M$ of the
reference particle measured from the pericenter, related to the mean motion $n$
by $M=nt$; in other words, $M/(2\pi)$ equals time in units of orbital periods.
The background Alfv\'en speed is $v_\su A\eqdef B/\rho^{1/2}$; from this we
derive the \definition{Alfv\'en parameter} $f^\su m_\mu\eqdef k_\mu v_\su A/n$,
which compares the frequencies of \acp{MHD} waves and mean orbital motion. The
\definition{acoustic parameter} $f^\su s_\mu\eqdef k_\mu c_\su s/n$ does the
same for sound waves. The other matrix elements are
\begin{alignat}{2}
\abbrev A(M)
  &\eqdef -2\Gamma^\lambda_{\phi\phi}\Omega/\lambda
  &&= 2\Omega\abbrev G, \\
\abbrev B(M)
  &\eqdef -\lambda(\Omega_\lambda+2\Gamma^\phi_{\lambda\phi}\Omega)
  &&= -\tfrac12\Omega, \\
\abbrev C(M)
  &\eqdef -(\Omega_\phi+2\Gamma^\phi_{\phi\phi}\Omega)
  &&= -2\Omega\abbrev H, \\
\abbrev D(M)
  &\eqdef \lambda\Omega_\lambda
  &&= -\tfrac32\Omega, \\
\abbrev E(M)
  &\eqdef \Omega_\phi
  &&= -2\Omega\abbrev H,
\end{alignat}
and
\begin{align}
\abbrev F^{\su s,\su m}_\lambda(M) &\eqdef
  -(g^{\lambda\lambda}if^{\su s,\su m}_\xi
  +\lambda g^{\lambda\phi}if^{\su s,\su m}_\eta), \\
\abbrev F^{\su s,\su m}_\phi(M) &\eqdef
  -(\lambda g^{\lambda\phi}if^{\su s,\su m}_\xi
  +\lambda^2g^{\phi\phi}if^{\su s,\su m}_\eta).
\end{align}
Here
\begin{alignat}{2}
\Omega(M) &\eqdef n^{-1}(\ods\theta t) &&= (1-e^2)^{-3/2}(1+e\cos\theta)^2, \\
\Omega_\lambda(M) &\eqdef n^{-1}\partial_\lambda(n\Omega)
  &&= -\tfrac32\Omega/\lambda, \\
\Omega_\phi(M) &\eqdef \partial_\phi\Omega &&= -2\Omega\abbrev H,
\end{alignat}
and
\begin{align}
\abbrev G(M) &\eqdef 1/(1+e\cos\theta), \\
\abbrev H(M) &\eqdef e\sin\theta/(1+e\cos\theta),
\end{align}
with $\theta(M)$ being the true anomaly of the reference particle. Expressions
for the inverse metric $g^{\mu\nu}$ and the Christoffel symbol
of the second kind $\Gamma^\mu_{\nu\rho}$ are found in Equations~(B13)--(B19)
of \citet{2014MNRAS.445.2621O}. Since the matrix elements are either constant
or $M$\nobreakdash-dependent with period $2\pi$ regardless of $e$, choosing $M$
as the time variable means that our results are independent of the semimajor
axis of the reference particle. In the circular limit, $\abbrev A$ is the
centrifugal force, $\abbrev B$ relates to the Coriolis force, $\abbrev D$
encodes orbital shear, and $(\Omega,\abbrev G,\abbrev H)=(1,1,0)$.

Circular \ac{MRI} originates from the destabilization of slow magnetosonic
waves in differentially rotating disks \citep{1998RvMP...70....1B}. Since these
waves are virtually incompressible, we, like \citet{1991ApJ...376..214B}, are
motivated to look firstly for similarly incompressible perturbations in
eccentric \ac{MRI}. Orbital shear creates nonzero horizontal components of the
velocity perturbation, so an incompressible perturbation must have a vertical
wavevector, that is, $\mathrelp\propto e^{ik_\zeta\zeta}$.

The adoption of a vertical wavevector means $k_\xi$, $k_\eta$, and $w^\zeta$
all vanish. The incompressible limit is characterized by $\Omega,f^\su m_\mu\ll
f^\su s_\mu$; when the wavevector is vertical, it is equivalent to setting $u$
and $v^\zeta$ to zero. Under these two assumptions, \cref{eq:solenoidal} is
automatically satisfied while \cref{eq:full MRI} simplifies significantly to
\begin{equation}\label{eq:shearing-box MRI}
\od{}M\begin{pmatrix} v^\xi \\ \lambda v^\eta \\ w^\xi \\ \lambda w^\eta
\end{pmatrix}=
\begin{pmatrix}
0 & \abbrev A & if & 0 \\
\abbrev B & \abbrev C & 0 & if \\
if & 0 & 0 & 0 \\
0 & if & \abbrev D & \abbrev E
\end{pmatrix}
\begin{pmatrix} v^\xi \\ \lambda v^\eta \\ w^\xi \\ \lambda w^\eta
\end{pmatrix},
\end{equation}
where $k\eqdef k_\zeta$ and $f\eqdef f^\su m_\zeta$ for brevity, and we assume
without loss of generality that $f\ge0$. The perturbation consists of the
two-dimensional velocity sector $(v^\xi,\lambda v^\eta)$ and the
two-dimensional magnetic sector $(w^\xi,\lambda w^\eta)$ in velocity units. The
matrix in \cref{eq:shearing-box MRI} splits into
\begin{equation}
\begin{pmatrix}
0 & \abbrev A & 0 & 0 \\
\abbrev B & \abbrev C & 0 & 0 \\
0 & 0 & 0 & 0 \\
0 & 0 & \abbrev D & \abbrev E
\end{pmatrix}+
\begin{pmatrix}
0 & 0 & if & 0 \\
0 & 0 & 0 & if \\
if & 0 & 0 & 0 \\
0 & if & 0 & 0
\end{pmatrix}.
\end{equation}
The first term is parametrized only by $e$ and the second term only by $f$;
therefore, the behavior of eccentric \ac{MRI} can be fully understood by
studying \cref{eq:shearing-box MRI} for all $(e,f)$. The first term is
time-dependent and describes how orbital variation excites oscillation within
each sector. The second term is time-independent, and describes how the
background magnetic field couples the two sectors and creates magnetic
oscillation. Both oscillations are themselves stable, but their coupling may
give rise to instability: for $e=0$, instability takes the form of circular
\ac{MRI} \citep{1991ApJ...376..214B}; for $e>0$, instability results from an
extension of circular \ac{MRI} to eccentric disks (\cref{sec:growth}) or the
parametric interaction between velocity and magnetic sectors
(\cref{sec:implications}).

\subsection{Linearized \texorpdfstring{\acsp*{MHD}}{MHD} equations in a
cylindrical coordinate basis}
\label{sec:cylindrical}

We define a cylindrical coordinate system $(R,\varphi,z)$ confocal with the
orbital coordinate system, and we equip each point with the cylindrical
coordinate basis alongside the shearing-box coordinate basis, as in
\cref{fig:coordinates}. The coordinate systems are related by
$(R,\varphi)=(\lambda\abbrev G,\phi)$, so contravariant components transform as
\begin{equation}\label{eq:coordinate transformation}
\begin{pmatrix} v^R \\ Rv^\varphi \end{pmatrix}=
\abbrev G\begin{pmatrix} 1 & \abbrev H \\ 0 & 1 \end{pmatrix}
\begin{pmatrix} v^\xi \\ \lambda v^\eta \end{pmatrix},
\end{equation}
and similarly for $(w^\xi,\lambda w^\eta)$. Useful properties of cylindrical
components are exposed when we convert \cref{eq:shearing-box MRI} to the
cylindrical coordinate basis using \cref{eq:coordinate transformation}:
\begin{equation}\label{eq:cylindrical MRI}
\od{}M\begin{pmatrix} v^R \\ Rv^\varphi \\ w^R \\ Rw^\varphi \end{pmatrix}=
\begin{pmatrix}
V_1 & V_2 & if & 0 \\
V_3 & V_4 & 0 & if \\
if & 0 & W_1 & W_2 \\
0 & if & W_3 & W_4
\end{pmatrix}
\begin{pmatrix} v^R \\ Rv^\varphi \\ w^R \\ Rw^\varphi \end{pmatrix},
\end{equation}
where
\begin{equation}\label{eq:submatrix}
\begin{pmatrix} V_1 & V_2 \\ V_3 & V_4 \end{pmatrix}\eqdef\tfrac12\Omega
\begin{pmatrix}
\abbrev H & (1-e^2)\abbrev G^2+3 \\ -1 & -\abbrev H
\end{pmatrix}
\end{equation}
and
\begin{equation}\label{eq:submatrix property}
\begin{pmatrix} V_1 & V_2 \\ V_3 & V_4 \end{pmatrix}
\begin{pmatrix} 0 & -1 \\ 1 & 0 \end{pmatrix}-
\begin{pmatrix} 0 & -1 \\ 1 & 0 \end{pmatrix}
\begin{pmatrix} W_1 & W_2 \\ W_3 & W_4 \end{pmatrix}=\tfrac12(1-e^2)^{-1/2}.
\end{equation}

If $f=0$, the velocity and magnetic sectors decouple, and $(v^R,Rv^\varphi)$
evolves independently of $(w^R,Rw^\varphi)$. If $f\ne0$, \cref{eq:cylindrical
MRI} admits solutions of the form
\begin{equation}\label{eq:solution form}
(v^R,Rv^\varphi,w^R,Rw^\varphi)=(\psi_1,\psi_2,ig\psi_2,-ig\psi_1),
\end{equation}
where $\psi_1(M)$ and $\psi_2(M)$ are complex-valued functions and $g$ is a
root of
\begin{equation}\label{eq:family}
g-g^{-1}=\tfrac12(1-e^2)^{-1/2}f^{-1}.
\end{equation}
The correctness of this solution is evident upon substituting
\cref{eq:submatrix property,eq:solution form,eq:family} into
\cref{eq:cylindrical MRI}. \Cref{eq:cylindrical MRI} can therefore be recast
into the equivalent form
\begin{equation}\label{eq:reduced MRI}
\od{}M\begin{pmatrix} v^R \\ Rv^\varphi \end{pmatrix}=
\begin{pmatrix} V_1 & V_2-fg \\ V_3+fg & V_4 \end{pmatrix}
\begin{pmatrix} v^R \\ Rv^\varphi \end{pmatrix}.
\end{equation}
Since the right-hand side of \cref{eq:family} is positive, its two roots
satisfy $g>1$ and $-1<g<0$ respectively. Two roots beget two solution families:
Positive\nobreakdash-$g$ solutions have magnetic components that are stronger
than velocity components, while negative\nobreakdash-$g$ solutions have the
opposite situation. The Maxwell stress is stronger than the Reynolds stress
when $g>0$, and it is the other way around when $g<0$. Increasing $e$ or
decreasing $f$ causes $\abs g$ to move further away from unity, enhancing the
contrast between velocity and magnetic components for both families. Because
$g$ can be readily inferred from $(e,f)$ and from the solution family, we shall
report only the time-evolution of $(v^R,Rv^\varphi)$.

\subsection{Floquet theory}
\label{sec:Floquet}

\Cref{eq:full MRI,eq:shearing-box MRI,eq:cylindrical MRI,eq:reduced MRI} of
eccentric \ac{MRI}, and \cref{eq:model matrix} of the toy model to be
introduced in \cref{sec:oscillator}, all have the form
\begin{equation}\label{eq:Floquet}
\od{\vec x}t=\mat A(t)\vec x(t),
\end{equation}
where $\vec x(t)$ is a vector and $\mat A(t)$ is a periodic matrix with period
$T$. For eccentric \ac{MRI}, $T$ is the orbital period. We cannot derive a
dispersion relation from this equation, so we turn to the theory of
\citet{1883ASENS..12...47F!}.

Consider the complex-valued equation
\begin{equation}\label{eq:Floquet matrix}
\od{\mat X}t=\mat A(t)\mat X(t),
\end{equation}
where $\mat X(t)$ is a matrix. A matrix-valued function $\mat F(t)$ is called a
\definition{fundamental matrix} if $\mat F(t)$ is a solution of
\cref{eq:Floquet matrix} and $\det\mat F(t)\ne0$ for all $t$. We can convince
ourselves that $\mat F(t)\mat C$, where $\mat C$ is a constant matrix, is a
fundamental matrix if and only if $\det\mat C\ne0$. In addition, $\mat F(t+T)$
is also a fundamental matrix.

The fundamental matrix $\mat G(t)$ satisfying $\mat G(0)=\mat1$ is called the
\definition{principal fundamental matrix}. Now $\mat G(t+T)$ and $\mat G(t)\mat
G(T)$ are both fundamental matrices with the same value at $t=0$, thus $\mat
G(t+T)=\mat G(t)\mat G(T)$ by the uniqueness of the solution; in other words,
the \definition{monodromy matrix} $\mat G(T)$ advances $\mat G(t)$ by a period.

Complex matrices, barring some exceptions such as nilpotent matrices, are
diagonalizable. Hence we set $\mat G(T)=\mat E\mat D\mat E\inv$, where $\mat
E\eqdef(\vec e_1,\vec e_2,\dotsc)$ is the matrix of column eigenvectors and
$\mat D\eqdef\diag(\alpha_1,\alpha_2,\dotsc)$ is the diagonal matrix of
eigenvalues, also called \definition{Floquet multipliers}. Consider the
fundamental matrix
\begin{equation}\label{eq:mode matrix definition}
\mat M(t)\eqdef\mat G(t)\mat E;
\end{equation}
clearly $\mat M(0)=\mat E$ and $\mat M(T)=\mat E\mat D$. This means if $\vec
x_j(t)$ is the $j$th column of $\mat M(t)$, then $\vec x_j(t)$ solves
\cref{eq:Floquet}, $\vec x_j(0)=\vec e_j$, and $\vec x_j(T)=\alpha_j\vec e_j$.

Our task in solving \cref{eq:Floquet} therefore reduces to finding $\mat G(T)$
by numerically integrating \cref{eq:Floquet matrix} over one period with the
identity matrix as the initial condition, and then computing the eigenvalues
and eigenvectors of $\mat G(T)$. Each eigenvector $\vec e_j$ produces one mode
$\vec x_j(t)$ of the full solution. Note that $\vec e_j$ and $\vec x_j(t)$ are
defined up to proportionality.

If we let $\mat P(t)\eqdef\mat M(t)\exp((-t/T)\ln\mat D)$, where $\exp$ and
$\ln$ are matrix exponentiation and logarithm respectively, then
\begin{align}
\nonumber \mat P(t+T) &= \mat M(t+T)\mat D\inv\exp((-t/T)\ln\mat D) \\
\nonumber &= \mat G(t)\mat G(T)\mat E\mat D\inv\exp((-t/T)\ln\mat D) \\
\nonumber &= \mat G(t)\mat E\exp((-t/T)\ln\mat D) \\
&= \mat P(t).
\end{align}
Thus we can write a mode as
\begin{equation}\label{eq:mode}
\vec x_j(t)\eqdef\alpha_j^{t/T}\vec p_j(t),
\end{equation}
where $\vec p_j(t)$ is periodic with period $T$. If $\mat A(t)$ in
\cref{eq:Floquet,eq:Floquet matrix} is constant, then $\mat P(t)$ is also
constant, so all modes are either exponential or sinusoidal, and their
respective growth rates or oscillation frequencies are given by the diagonal of
$\ln\mat D/T$. Note that $\vec x_j(t)$ is periodic if and only if $\alpha_j$ is
a root of unity or $\vec p_j(t)$ is constant.

A mode $\vec x_j(t)$ is \definition{stable} if $\abs{\alpha_j}\le1$ and
\definition{unstable} if $\abs{\alpha_j}>1$. The stability of \cref{eq:Floquet}
depends only on the mode with the largest $\abs{\alpha_j}$; we call this mode
the \definition{most unstable mode}, and let the \definition{growth per period}
of \cref{eq:Floquet} be $\gamma\eqdef\max_j\ln\abs{\alpha_j}$.

Since any fundamental matrix $\mat F(t)$ is a solution of \cref{eq:Floquet
matrix}, we have $\ods{(\ln\det\mat F)}t=\tr\mat A(t)$; in particular,
\begin{equation}\label{eq:eigenvalue product}
\alpha_1\alpha_2\dotsm=\det\mat G(T)=\exp\int_0^Tdt\,\tr\mat A(t).
\end{equation}
If the integral vanishes, $\alpha_1\alpha_2\dotsm=1$ and so $\gamma\ge0$.

\subsection{Application of Floquet theory to eccentric
\texorpdfstring{\acs*{MRI}}{MRI}}
\label{sec:applied Floquet}

The special nature of the $2\times2$ matrix $\mat A(t)$ in \cref{eq:reduced
MRI,eq:model matrix} leads to additional useful properties. Since $\mat A(t)$
is real, $\mat G(T)$ and its trace, $\alpha_1+\alpha_2$, are both real; since
$\mat A(t)$ is traceless, \cref{eq:eigenvalue product} yields
$\alpha_1\alpha_2=1$. Thus either $\abs{\alpha_1}=\abs{\alpha_2}=1$, which
gives two stable modes; or $\alpha_1$ and $\alpha_2$ are both real, which gives
one stable mode and one unstable mode.

Consider the latter case. If $\alpha_1,\alpha_2>0$, then both modes are
\definition{sign-preserving} in the sense that each component of $\vec x_j(t)$
retains the same sign after a period; conversely, if $\alpha_1,\alpha_2<0$,
then the two modes are \definition{sign-reversing} because each component of
$\vec x_j(t)$ flips sign. Moreover, \cref{eq:mode matrix definition} implies
$\mat M(T)=\mat G(T)\mat E$, the $j$th column of which is $\alpha_j\vec
e_j=\mat G(T)\vec e_j$. If $\mat G(T)$ is diagonal, then $\mat E$ is the
identity matrix and the components of $\vec e_j$ are real. If $\mat G(T)$ is
not diagonal, then the real matrix $\mat G(T)$ mixes the two components of
$\vec e_j$ to give $\vec e_j$ times a real scalar $\alpha_j$; this can only be
so if the components share the same complex phase. We can therefore take $\vec
x_j(0)=\vec e_j$ to be real without loss of generality; \cref{eq:reduced
MRI,eq:model matrix} then compel $\vec x_j(t)$ to be real for all $t$.

The results in the previous paragraphs have important implications for
eccentric \ac{MRI}. \Cref{eq:shearing-box MRI,eq:cylindrical MRI} have four
modes each, dividing into pairs of two: a pair from solving \cref{eq:reduced
MRI} with $g>0$, another pair from solving the same equation with $g<0$
(\cref{sec:cylindrical}). Here we showed that each pair comprises either two
stable modes, or one stable mode and one unstable mode, thus eccentric \ac{MRI}
can have at most two unstable modes, one for each sign of $g$. Furthermore,
while we can choose $(v^R,Rv^\varphi)$ in \cref{eq:reduced MRI} to be real at
all times, \cref{eq:solution form} simultaneously makes $(w^R,Rw^\varphi)$
imaginary, so the mode is restricted to the velocity sector. Since the
perturbation is $\mathrelp\propto e^{ik\zeta}$ (\cref{sec:MRI}), selecting a
different complex phase means observing at a different height above the
reference particle. If the velocity sector is real and the magnetic sector
imaginary at some $\zeta$, then the velocity sector is imaginary and the
magnetic sector real at $\zeta+\tfrac12j\pi k^{-1}$ for any odd integer $j$.

\section{Eccentric MRI}
\label{sec:results}

\subsection{Growth per orbit of the most unstable modes}
\label{sec:growth}

\begin{figure}
\includegraphics{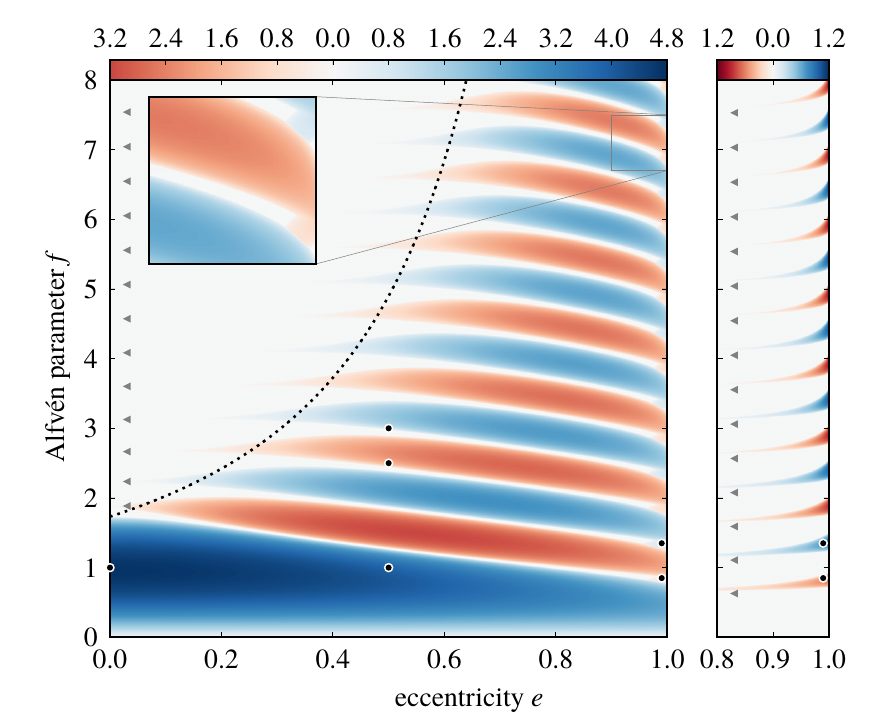}
\caption{Plot of $\gamma$ for eccentric \ac{MRI} as a function of eccentricity
$e$ and Alfv\'en parameter $f$. Here $e^\gamma\ge1$ is the absolute value of
the amplification of the most unstable mode per orbit (\cref{sec:Floquet}),
hence colored regions are unstable; blue indicates sign-preserving regions and
red indicates sign-reversing regions (\cref{sec:applied Floquet}). Dots mark
six $(e,f)$ chosen for closer examination (\cref{sec:phase,sec:stress}).
\textit{Left panel:} The classical band is the lowermost horizontal unstable
region, parametric bands are the horizontal unstable regions above it, and
horns are the unstable regions peeking out between adjacent parametric bands,
as highlighted by the inset (\cref{sec:growth}). Parametric bands sharply
narrow as $e\to0$, becoming points at $e=0$; the points, marked by gray
triangles, fall at values of $f$ for which stable circular band modes complete
integer or half-integer numbers of oscillations per orbit
(\cref{sec:implications}). The dotted curve $(1-e)^3f^2=3$ is the threshold
between the mostly stable regime to the left and the mostly unstable regime to
the right (\cref{sec:implications}). \textit{Right panel:} Cutout of the left
panel at $e\ge0.8$, showing only horns (\cref{sec:growth}). Gray triangles mark
the points at $e=0$ to which horns taper (\cref{sec:implications}).}
\label{fig:MRI growth}
\end{figure}

\Cref{eq:shearing-box MRI,eq:cylindrical MRI,eq:reduced MRI} are numerically
integrated over one orbit, and the growth per orbit $\gamma$ of eccentric
\ac{MRI} is given by the eigenvalue of the monodromy matrix with the greatest
complex magnitude (\cref{sec:Floquet}). The left panel of \cref{fig:MRI growth}
displays $\gamma$ as a function of $(e,f)$; \definition{unstable regions} are
where $\gamma>0$. Circular disks correspond to $e=0$; in agreement with
\citet{1991ApJ...376..214B}, we find \ac{MRI} if $0<f^2<3$, that is, if the
background magnetic field is weak and the wavenumber is small, and stability
otherwise.

The behavior for arbitrary $(e,f)$ is more complicated. \definition{Bands} and
\definition{horns} are respectively unstable regions found by solving
\cref{eq:reduced MRI} with positive and negative values of $g$ that satisfy
\cref{eq:family}; bands further divide into the \definition{classical band} and
\definition{parametric bands}.

The classical band is the extension of the unstable region of circular \ac{MRI}
to $e>0$. It contains positive\nobreakdash-$g$ unstable modes, whose magnetic
components are larger than their velocity components (\cref{sec:cylindrical}).
Growth is fastest at $(e,f)=(0,\tfrac14\smash{\sqrt{15}})$, with
$\gamma=\tfrac32\pi$. The width of the classical band, as measured in the
$f$\nobreakdash-direction, and its $\gamma$ at fixed $f$ both fall by a factor
of \num{\approx2} from $e=0$ to $e=1$; this is because these modes grow with
the help of constant orbital shear just as in circular \ac{MRI}, but when $e$
is large, orbital shear, encapsulated by $\lambda\Omega_\lambda$ and
$\Omega_\phi$, is small during the long time spent near apocenter.

Parametric bands contain positive\nobreakdash-$g$ unstable modes not included
in the classical band; like classical-band modes, the magnetic components of
these modes are larger than their velocity components (\cref{sec:cylindrical}).
Parametric bands appear as banana-shaped unstable regions above the classical
band that are, loosely speaking, elongated in the $e$\nobreakdash-direction and
stacked in the $f$\nobreakdash-direction. Overall, $\gamma$ in parametric bands
is about half the largest $\gamma$ for circular \ac{MRI}; more precisely,
bandwidth and $\gamma$ both increase with $e$ at fixed $f$ up to a broad
maximum, then decrease slowly toward $e=1$, while $\gamma$ decreases slightly
with increasing $f$ at fixed $e$. Parametric bands are spaced at
$f$\nobreakdash-steps of $\mathrelp\approx\tfrac12$, with adjacent parametric
bands separated by a narrow but finite gap for all $e$, and they alternate
between sign-preserving and sign-reversing in the $f$\nobreakdash-direction
(\cref{sec:applied Floquet}). The regular spacing of parametric bands, their
clean separation from one another, and their narrowing toward $e=0$ all suggest
an origin related to parametric resonance (\cref{sec:implications}).

Horns contain negative\nobreakdash-$g$ unstable modes, whose velocity
components are larger than their magnetic components (\cref{sec:cylindrical}).
The right panel of \cref{fig:MRI growth} depicts horns in isolation. Their
width and $\gamma$ both increase with $e$, attaining noticeable width only at
$e\gtrsim0.8$; they are also regularly spaced in $f$. Because horns have
smaller $\gamma$ than bands, they are mostly buried underneath bands in the
left panel, which portrays only the most unstable modes of
\cref{eq:shearing-box MRI,eq:cylindrical MRI}; however, horns do emerge between
bands when bandgaps widen at $e\gtrsim0.95$. At such high $e$, horns have
$\gamma$ approximately half that of bands, so horn modes can be as important as
band modes in stirring \ac{MHD} turbulence.

The largest $\gamma$ at any given $e>0$ is generally a factor of a few smaller
than that at $e=0$. However, the largest $f$ that permits instability rises
rapidly with $e$ from the circular-limit value of $\smash{\sqrt3}$, so the
$f$\nobreakdash-range over which \ac{MRI} operates is substantially wider at
$e>0$. Horns also tend to fill in bandgaps at $e\gtrsim0.95$, making more
values of $f$ susceptible to \ac{MRI} at high $e$.

While there is no exponential growth at band and horn edges because $\gamma=0$
there by definition, growth in general may still occur (\cref{sec:edge}).

\subsection{Limiting behavior of modes}
\label{sec:limit}

The division of eccentric \ac{MRI} modes into band and horn modes has physical
significance, which is most easily appreciated in the $e=0$ and $f=0$ limits.

The $e=0$ limit reproduces circular \ac{MRI}; these \definition{circular modes}
follow the dispersion relation \citep[e.g.,][]{1991ApJ...376..214B}
\begin{equation}\label{eq:dispersion}
(\omega/n)^2=f^2+\tfrac12\mp(4f^2+\tfrac14)^{1/2}.
\end{equation}
The upper sign yields $(\omega/n)^2<0$ if and only if $0<f^2<3$, while the
lower sign has $(\omega/n)^2\ge1$ for all $f$. Since the $0<f^2<3$ part of the
$f$\nobreakdash-axis in \cref{fig:MRI growth} is covered only by the classical
band, we associate the upper and lower signs with bands and horns respectively.
Band modes with $f^2\ge3$ and all horn modes are stable at $e=0$; these stable
circular modes are destabilized by orbital variation through parametric
resonance at $0<e\ll1$, producing parametric bands and horns respectively
(\cref{sec:implications}).

The $f=0$ limit is trickier. \Cref{eq:reduced MRI} does not apply, so we cannot
classify modes as band or horn. Moreover, \cref{eq:shearing-box
MRI,eq:cylindrical MRI} have non-diagonalizable monodromy matrices, leaving us
with just the three modes given in \cref{sec:edge}. All three modes are stable
because the magnetic and velocity sectors decouple (\cref{sec:cylindrical}) and
the two sectors are individually stable (\cref{sec:MRI}). The first two modes
have vanishing $(w^\xi,\lambda w^\eta)$ and periodic $(v^\xi,\lambda v^\eta)$,
hence we identify them as epicycles, or inertial modes with vertical
wavevectors. The third mode has vanishing $(v^\xi,\lambda v^\eta)$,
corresponding to the situation where a gas packet is displaced along the orbit
without any change in velocity; we call this neutrally stable mode a
\definition{sliding mode}, and it is analogous to the azimuthal displacements
in circular disks discussed by \citet{1991ApJ...376..214B}. The magnetic field
perturbation of this mode is frozen into the background flow, and $\lambda
w^\eta$ varies periodically in proportion to the orbital speed.

Although inertial and sliding modes are, strictly speaking, neither band nor
horn, we can associate them with band and horn modes at $f>0$ by studying how
these latter modes behave as $f\to0$. We find that the two band modes merge to
the sliding mode, whereas the two horn modes tend independently toward the two
inertial modes. Just as azimuthal displacements in circular disks are readily
destabilized by orbital shear in the presence of a weak magnetic field
\citep{1991ApJ...376..214B}, the sliding mode is destabilized at $0<f\ll1$ to
produce the classical band. Horn modes are more closely related to epicycles,
which are stable in Keplerian disks \citep[e.g.,][]{1917RSPA...93..148R!};
consequently, horn modes are not destabilized at $0<f\ll1$, and are
destabilized to any appreciable extent only at $e\gtrsim0.8$.

\subsection{Time-evolution of unstable modes}
\label{sec:phase}

When $e>0$, the matrices in \cref{eq:full MRI,eq:shearing-box
MRI,eq:cylindrical MRI,eq:reduced MRI} are time-dependent, hence unstable modes
do not grow at a steady exponential rate, nor do their components bear a
constant ratio; instead, components vary at different paces in the course of an
orbit, in such a way that they are all multiplied by a common factor after a
complete orbit, as guaranteed by \cref{eq:mode}. It is therefore instructive to
examine in detail how unstable modes evolve within a single orbit.

\begin{figure}
\includegraphics{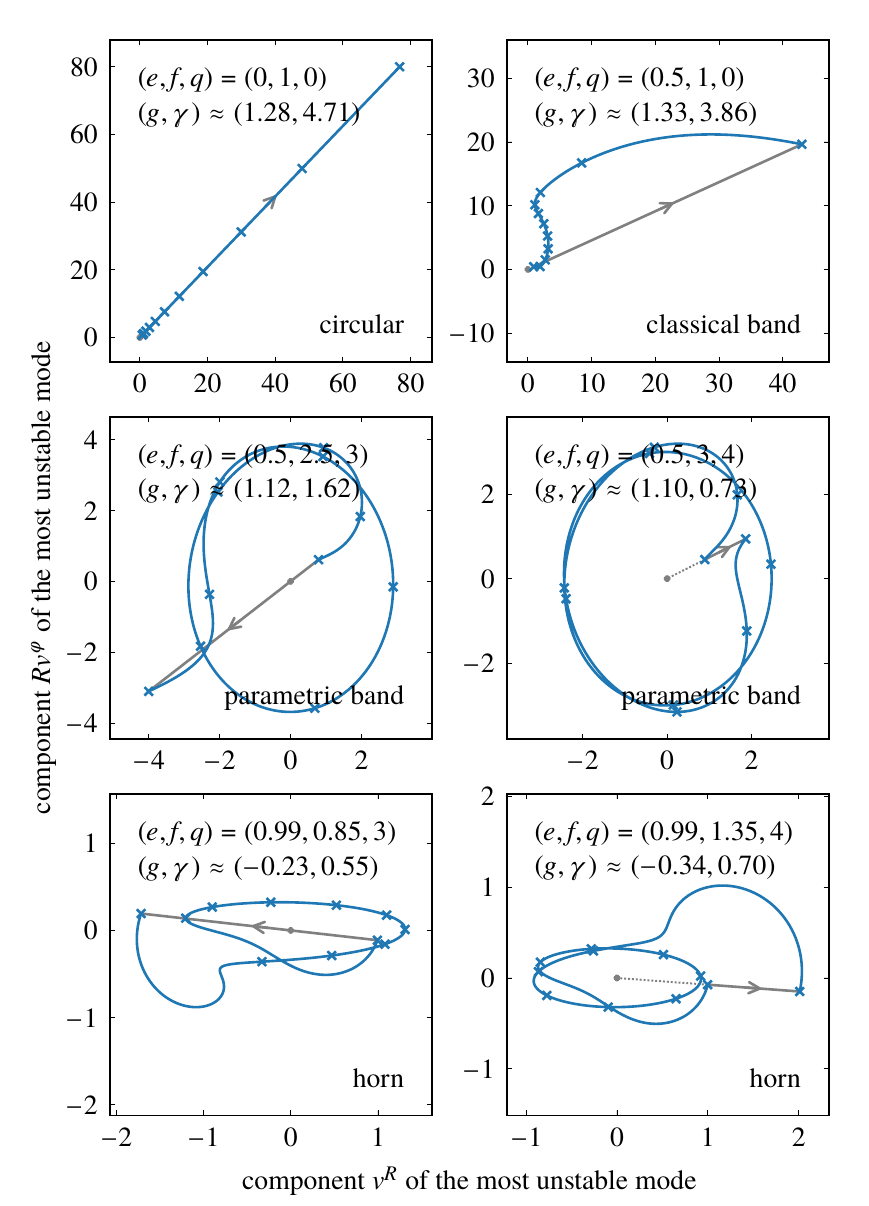}
\caption{Pericenter-to-pericenter trajectory of $(v^R,Rv^\varphi)$ of the most
unstable mode of eccentric \ac{MRI} for the six $(e,f)$ indicated by dots in
\cref{fig:MRI growth}. Band and horn modes have $g>0$ and $g<0$ respectively
(\cref{sec:cylindrical}). Trajectories are marked every tenth of an orbital
period with a cross. The origin, shown as a dot, lies on the solid gray line
connecting the beginning of a trajectory to its end; the end is $e^\gamma>1$
times as distant from the origin as the beginning.}
\label{fig:phase}
\end{figure}

To accentuate the difference between bands and horns, we pick six $(e,f)$ from
where they do not overlap, that is, where precisely one mode is unstable
(\cref{sec:applied Floquet}); our selection is indicated by dots in
\cref{fig:MRI growth}. The six unstable modes include a circular band mode, a
classical-band mode, two modes from adjacent parametric bands, and two modes
from adjacent horns. To determine the time-dependence of each mode, we
numerically integrate \cref{eq:reduced MRI} over an orbit with the pericenter
value of the mode as the initial condition; \cref{fig:phase} plots the
resulting trajectory of two components $(v^R,Rv^\varphi)$ of the mode. We
choose the complex phase of the perturbation such that $(v^R,Rv^\varphi)$ is
always real (\cref{sec:applied Floquet}); this is done purely for ease of
visualization and has no physical significance. \Cref{eq:solution form} takes
us from $(v^R,Rv^\varphi)$ to $(w^R,Rw^\varphi)$, which in this case is purely
imaginary. For sign-preserving modes (\cref{sec:applied Floquet}), the
trajectory of the subsequent orbit traces out the same shape magnified by a
factor of $e^\gamma>1$; for sign-reversing modes, the magnification is
$-e^\gamma<-1$, that is, the trajectory is enlarged and inverted with respect
to the origin.

For the circular band mode at $(e,f)=(0,1)$ in the top-left panel, the mode
grows exponentially, the trajectory is straight, and the same growth rate
applies to all components of the perturbation. For the classical-band mode at
$(e,f)=(0.5,1)$ in the top-right panel, however, orbital variation bends the
trajectory away from a straight line; this is symptomatic of the uneven growth
of different components within an orbit, and is a generic feature of \ac{MRI}
growth in eccentric disks.

For the parametric-band modes at $(e,f)=(0.5,2.5)$ and $(e,f)=(0.5,3)$ in the
center panels, the middle part of each trajectory, traversed while the gas
travels out to the apocenter and back, loops counterclockwise around the
origin; the mode does not grow along the loop, as evidenced by the confinement
of $(v^R)^2+(Rv^\varphi)^2$ to a finite range. The ends, corresponding to
pericenter passage, deviate from the loop; the deviation is outward whenever
the mode grows and inward whenever the mode decays. Although most growth takes
place near pericenter, $(v^R)^2+(Rv^\varphi)^2$ does not necessarily increase
monotonically throughout pericenter passage. For the $f=2.5$ mode, the
trajectory makes $\tfrac32$ turns around the origin; for the $f=3$ mode, the
trajectory goes around twice. As we discuss below, this \definition{winding
number} is always integer or half-integer.

For the horn modes at $(e,f)=(0.99,0.85)$ and $(e,f)=(0.99,1.35)$ in the bottom
panels, the trajectories are qualitatively the same as parametric-band modes,
except that the apocentric loop is clockwise and $\gamma$ is generally smaller.

Taking appropriate limits in \cref{eq:reduced MRI} yields physical insight
about the apocentric loop. Near apocenter, the diagonal elements of the matrix
in \cref{eq:reduced MRI} are small because $\abbrev H\approx0$. If additionally
\begin{equation}\label{eq:loop condition}
(V_2-fg)(V_3+fg)<0,
\end{equation}
\cref{eq:reduced MRI} describes stable oscillation in which $(v^R,Rv^\varphi)$
loosely traces out an ellipse of horizontal-to-vertical axis ratio
$[-(V_2-fg)/(V_3+fg)]^{1/2}$; note that $V_2>0$ and $V_3<0$. Horn modes always
satisfy \cref{eq:loop condition} because their $g<0$ (\cref{sec:cylindrical});
for them, magnetic tension and orbital forces drive oscillation together. Band
modes are harder to handle because \cref{eq:loop condition} is true only for
parts of the orbit; for mathematical expedience, we consider \cref{eq:loop
condition} only at apocenter, trusting that if it holds at apocenter, then the
continuity of $(V_2-fg)(V_3+fg)$ over $\theta$ would ensure it holds over a
finite range around apocenter as well. Band modes have $g>0$
(\cref{sec:cylindrical}), thus
$-V_3=\tfrac12\Omega=\tfrac12(1-e^2)^{-3/2}(1-e)^2\le\tfrac12(1-e^2)^{-1/2}=f(g-g^{-1})<fg$,
where we used \cref{eq:family} in the fourth step. It follows that
\cref{eq:loop condition} is equivalent to
$fg>V_2=\tfrac12(1-e^2)^{-1/2}+\tfrac32\Omega$; in other words, oscillation
occurs if magnetic tension beats orbital forces. The solution of the last
inequality is $f^2>\tfrac32(1+e)^{-3}(2-e)$, which includes all parametric
bands and part of the classical band, so modes there exhibit apocentric loops.
Furthermore, \cref{eq:family} yields $[\pds{(fg)}f]_e=2g/(1+g^2)>0$, hence when
$f$ is larger, $(V_2-fg)(V_3+fg)$ at apocenter is more negative, \cref{eq:loop
condition} is satisfied over a larger fraction of the orbit around apocenter,
and the trajectory spends more time looping near apocenter and less time
growing near pericenter. This is exactly the trend suggested by the three modes
at $e=0.5$ in \cref{fig:phase}.

The argument in the previous paragraph explains why the apocentric loop exists
and why the ends deviate from the apocentric loop. However, it is not very
useful near pericenter: For bands, the argument is frustrated by the fact that
the signs of $V_2-fg$ and $V_3+fg$ depend on $(e,f)$; for horns, the argument
hardly matters because their unstable modes appear only at $e\gtrsim0.8$
(\cref{sec:growth}), thus little time is spent where $\abbrev H\approx0$.
Instead, we advance another argument applicable to the near-pericenter
evolution of modes along the midlines of parametric bands and horns. Because
$\gamma$ reaches a local maximum there, we can reasonably expect that both ends
would be growing, which simplifies our considerations; the behavior along the
midline is also likely characteristic of the entire parametric band or horn.
Growth of midline modes is concentrated near pericenter because orbital shear
is necessary to draw out magnetic field perturbations, and orbital shear is the
strongest relative to magnetic tension in that part of the orbit. The time an
orbit spends near pericenter is $\mathrelp\sim[\Omega(\theta=0)]^{-1}$, and the
instantaneous growth rate is roughly the orbital shear, that is,
$\mathrelp\sim\Omega(\theta=0)$; their product is therefore always of order
unity, which may explain why $\gamma$ varies weakly with $e$ in \cref{fig:MRI
growth}.

The fact that unstable modes grow by a real factor every orbit
(\cref{sec:applied Floquet}) means that $(v^R,Rv^\varphi)$ at all pericenters
must lie on a line that includes the origin, as in \cref{fig:phase}.
Sign-preserving modes have integer winding numbers because $(v^R,Rv^\varphi)$
at successive pericenters are on the same side of the origin; sign-reversing
modes have half-integer winding numbers because $(v^R,Rv^\varphi)$ switches
sides every orbit (\cref{sec:applied Floquet}). Each band or horn has a single
winding number; we call twice this number its \definition{order} $q$. The
classical band has order $q=0$. The lowermost parametric band and horn have
orders $q=1$ and $q=3$ respectively; each band or horn above is one order
higher. The winding number increases with $f$ because stronger magnetic tension
drives faster oscillation around the apocentric loop.

\subsection{Angular momentum and energy transport by unstable modes}
\label{sec:stress}

Denote the shearing-box and cylindrical coordinate bases by $(\uvec
e_\lambda,\uvec e_\phi)$ and $(\uvec e_R,\uvec e_\varphi)$ respectively, and
their normalized versions by $(\uvec e_{\hat\lambda},\uvec e_{\hat\phi})$ and
$(\uvec e_{\hat R},\uvec e_{\hat\varphi})$; the normalized coordinate bases are
depicted in the top half of \cref{fig:coordinates}. Recall that in differential
geometry, a coordinate basis is defined to be tangent to the coordinate curves;
specifically, $\uvec e_\lambda$ is tangent to curves of constant $\phi$ and
$z$, while $\uvec e_R$ is tangent to curves of constant $\varphi$ and $z$.
Since these two sets of curves coincide, we have $\uvec e_\lambda\parallel\uvec
e_R$ and $\uvec e_{\hat\lambda}=\uvec e_{\hat R}$.

\begin{figure}
\includegraphics{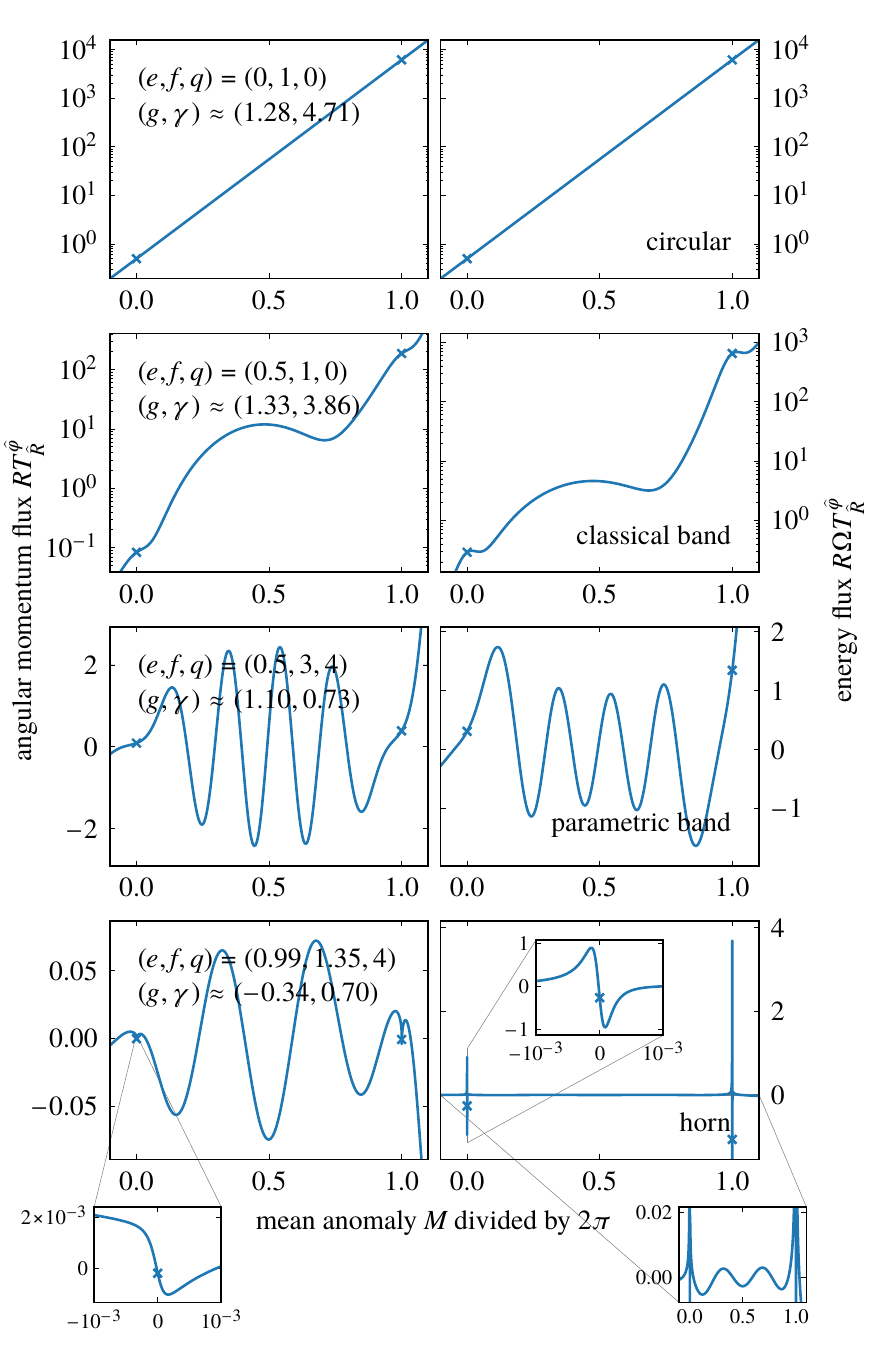}
\caption{Radial angular momentum (left) and energy (right) fluxes due to the
Reynolds stress associated with the most unstable mode of eccentric \ac{MRI}
for four of the six $(e,f)$ indicated by dots in \cref{fig:MRI growth}. Band
and horn modes have $g>0$ and $g<0$ respectively (\cref{sec:cylindrical}).
Perturbations are normalized such that $(v^R)^2+(Rv^\varphi)^2=1$ at $M=0$.
Crosses mark pericenter fluxes at $M/(2\pi)\in\{0,1\}$; the right cross is at a
flux level $e^{2\gamma}>1$ times the left cross. The top two rows have
logarithmic vertical scales, while the bottom two rows have linear vertical
scales.}
\label{fig:stress}
\end{figure}

The $\hat\varphi\hat R$\nobreakdash-component of the Reynolds stress tensor is
$\strcyl=v^RRv^\varphi$; the associated angular momentum and energy fluxes are
$R\strcyl$ and $R\Omega\strcyl$ respectively. \Cref{fig:stress} plots the two
fluxes over an orbit for a circular band mode, a classical-band mode, a
parametric-band mode, and a horn mode. Fluxes in the subsequent orbit have the
same shape, but the overall normalization is $e^{2\gamma}>1$ times greater
(\cref{sec:Floquet}). Since $(v^R,Rv^\varphi)$ is chosen to be real
(\cref{sec:applied Floquet}), \cref{fig:stress} shows fluxes at $\zeta$ such
that the Reynolds stress is the greatest. The Maxwell stress reaches its
maximum at a different $\zeta$ (\cref{sec:applied Floquet}); the Maxwell stress
there is $g^2$ times the Reynolds stress here, where $\abs g>1$ for band modes
and $\abs g<1$ for horn modes (\cref{sec:cylindrical}).

The $\hat\varphi$\nobreakdash-momentum flux in the
$\hat\lambda$\nobreakdash-direction $\strorb$ is obtained by performing a
coordinate transformation from cylindrical to shearing-box on the lower index
of $\strcyl$; lower indices transform covariantly as the basis, thus
$\strorb=\strcyl$. Intuitively, $\strorb$ is the flux of angular momentum
through the elliptical, constant\nobreakdash-$\lambda$ line segment in the
bottom half of \cref{fig:coordinates}, and $\strcyl$ is the same through the
circular, constant\nobreakdash-$R$ line segment. The greater length of the
former line segment is made up for by its obliquity to the latter, hence the
two fluxes are the same.

For the circular band mode at $(e,f)=(0,1)$ in the first row, fluxes grow
exponentially, yielding straight lines on semi-logarithmic plots. For the
classical-band mode at $(e,f)=(0.5,1)$ in the second row, orbital variation
bends the trajectory of $(v^R,Rv^\varphi)$ away from exponential growth
(\cref{sec:phase}), so fluxes do not increase monotonically. Note that not all
classical-band modes have $\strcyl>0$ throughout the orbit. If $e$ and $f$ are
both large, $v^R$ switches sign and then switches back pre-pericenter; if $e$
is large but $f$ is small, $Rv^\varphi$ changes sign in like manner
post-pericenter. In either case, $\strcyl<0$ over a fraction of the orbit.

For the parametric-band mode at $(e,f)=(0.5,3)$ in the third row, and for the
horn mode at $(e,f)=(0.99,1.35)$ in the fourth row, the trajectory of
$(v^R,Rv^\varphi)$ makes a roughly elliptical loop around the origin near
apocenter (\cref{sec:phase}); $v^RRv^\varphi$ therefore oscillates almost
sinusoidally over that part of the orbit, and $\strcyl$ is also more or less
sinusoidal considering that $R$ varies slowly there. The sign change of
$\strcyl$ means that stresses sometimes move angular momentum and energy
outward, and sometimes inward; the sinusoidal nature of $\strcyl$ near
apocenter leads to strong cancellation between outward and inward fluxes.
Because the modes in \cref{fig:stress} have relatively small $f$, cancellation
may not be very conspicuous; at large $f$ however, where the apocentric loop
covers more of the orbit and the winding number is large (\cref{sec:phase}),
$\strcyl$ goes through many periods of sinusoidal oscillation near apocenter,
so we anticipate close to complete cancellation. Near pericenter, the mode
grows, and $\strcyl$ can be larger post-pericenter than pre-pericenter or the
other way around; this asymmetry can create net transport.

Parametric-band and horn modes fail to grow (\cref{sec:phase}) and $\strcyl$
integrates to a vanishing value over the same part of the orbit, namely, the
apocentric loop. The concurrence is unsurprising. Orbital shear feeds the
perturbation by draining energy from the background flow and converting it to
the kinetic and magnetic energy of the perturbation; in doing so, orbital shear
establishes a positive correlation between $v^R$ and $Rv^\varphi$, which leads
to net outward transport. Along the apocentric loop, orbital shear is too weak
compared to magnetic tension to do either.

\begin{figure}
\includegraphics{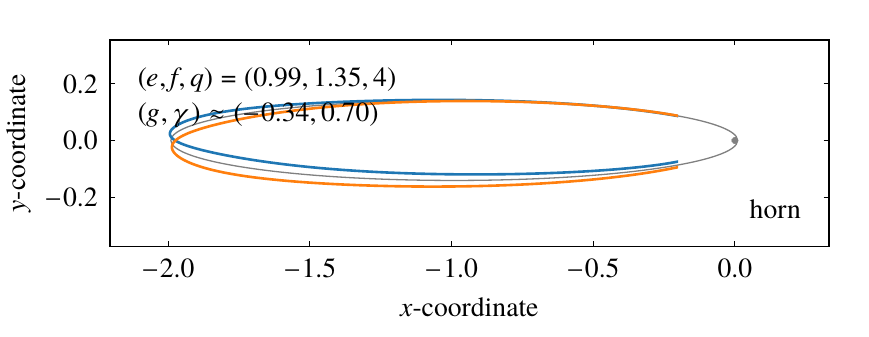}
\caption{Schematic illustration of differential apsidal precession due to
\ac{MHD} stresses. The orbital plane of the blue and orange particles is
described by a Cartesian coordinate system $(x,y)$ whose origin is at the point
mass. Stresses associated with the most unstable mode of eccentric \ac{MRI}
transfer momentum from the orange particle to the blue at a rate
$\mathrelp\propto\strcyl\,\uvec e_\varphi$ (\cref{fig:stress}); as a result,
their trajectories deviate from the gray Keplerian orbit in opposite senses.
These deviations, assumed small, grow by a factor of $e^{2\gamma}>1$ per orbit;
here they are exaggerated for clarity.}
\label{fig:precession}
\end{figure}

Net transport changes osculating orbital elements across the disk. When
$e\approx1$, a small increase in the argument of pericenter at one edge of the
disk and a corresponding decrease at the other leads to differential apsidal
precession and apocentric stream crossing as seen in \cref{fig:precession}.
This phenomenon is distinct from the differential apsidal precession described
by \citet{2001MNRAS.325..231O}, which requires a radial pressure gradient.

\section{Toy model}
\label{sec:model}

The complexity of \cref{eq:shearing-box MRI,eq:cylindrical MRI,eq:reduced MRI}
suggests that we may gain further insight from studying simpler versions of
them, ones stripped down to their essential elements. We discuss such a toy
model in this section.

\subsection{Frequency-modulated oscillator}
\label{sec:oscillator}

The physics of eccentric \ac{MRI} boils down to the interaction between
magnetic and orbital forces: The background magnetic field controls the
oscillation between velocity and magnetic sectors (\cref{sec:MRI}), while
orbital variation modulates the strength of this oscillation. We can expose
this interaction by eliminating $w^\xi$ and $\lambda w^\eta$ from
\cref{eq:shearing-box MRI}:
\ifapj
  \begin{equation}\label{eq:MRI oscillator}
  \odd{}M\begin{pmatrix} v^\xi \\ \lambda v^\eta \end{pmatrix}
  -\begin{pmatrix}
  0 & \abbrev A \\ \abbrev B+\abbrev D & \abbrev C+\abbrev E
  \end{pmatrix}\od{}M\begin{pmatrix} v^\xi \\ \lambda v^\eta \end{pmatrix}
  +\begin{pmatrix}
  f^2 & -\dot{\abbrev A} \\ \abbrev B\abbrev E-\dot{\abbrev B} & \abbrev J
  \end{pmatrix}
  \begin{pmatrix} v^\xi \\ \lambda v^\eta \end{pmatrix}=0,
  \end{equation}
\else
  \begin{multline}\label{eq:MRI oscillator}
  \odd{}M\begin{pmatrix} v^\xi \\ \lambda v^\eta \end{pmatrix}
  -\begin{pmatrix}
  0 & \abbrev A \\ \abbrev B+\abbrev D & \abbrev C+\abbrev E
  \end{pmatrix}\od{}M\begin{pmatrix} v^\xi \\ \lambda v^\eta \end{pmatrix} \\
  +\begin{pmatrix}
  f^2 & -\dot{\abbrev A} \\ \abbrev B\abbrev E-\dot{\abbrev B} & \abbrev J
  \end{pmatrix}
  \begin{pmatrix} v^\xi \\ \lambda v^\eta \end{pmatrix}=0,
  \end{multline}
\fi
where $\abbrev J(M)\eqdef f^2+\abbrev A\abbrev D+\abbrev C\abbrev
E-\dot{\abbrev C}$, and overdot denotes differentiation with respect to $M$.
This equation reduces to Equations~(106) and~(107) of
\citet{1998RvMP...70....1B} in the circular limit. The equation describes a
pair of coupled, damped oscillators; in the circular limit, the natural
frequencies of the two oscillators are $f$ and $(f^2-3)^{1/2}$ respectively,
but eccentric orbital motion causes periodic modulation of the latter natural
frequency.

\begin{figure}
\includegraphics{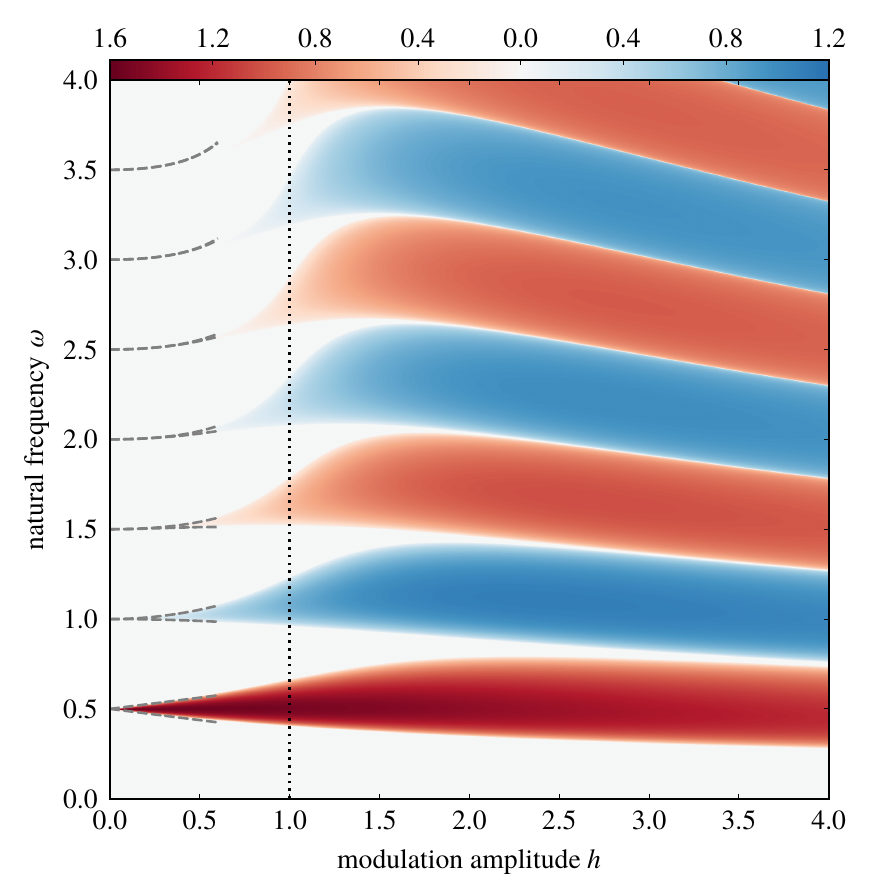}
\caption{Plot of $\gamma/(h\omega)$ for the toy model as a function of the
amplitude $h$ of frequency modulation and natural frequency $\omega$ in
\cref{eq:model}. Here $e^\gamma\ge1$ is the absolute value of the amplification
of the unstable mode per orbit (\cref{sec:Floquet}), hence colored regions are
unstable; blue indicates sign-preserving regions and red indicates
sign-reversing regions (\cref{sec:applied Floquet}). The dotted line is the
threshold between the mostly stable regime to the left and the mostly unstable
regime to the right. Dashed curves mark the edges of unstable regions computed
using perturbative methods (\cref{sec:parametric resonance}).}
\label{fig:model growth}
\end{figure}

This observation motivates the study of the toy model
\begin{equation}\label{eq:model}
\odd xt+\omega^2(1+h\cos t)x=0
\end{equation}
as a step toward better understanding eccentric \ac{MRI}. This equation governs
an oscillator whose frequency is periodically modulated around $\omega$, thus
$h$ and $\omega$ of the toy model are respectively analogous to $e$ and $f$ of
eccentric \ac{MRI}. The equation can be rewritten as
\begin{equation}\label{eq:model matrix}
\od{}t\begin{pmatrix} x \\ \dot x \end{pmatrix}=
  \begin{pmatrix} 0 & 1 \\ -\omega^2(1+h\cos t) & 0 \end{pmatrix}
  \begin{pmatrix} x \\ \dot x \end{pmatrix},
\end{equation}
where $\dot x\eqdef\ods xt$; analyzing this equation with the method developed
in \cref{sec:Floquet} results in \cref{fig:model growth}.

\Cref{fig:model growth} resembles \cref{fig:MRI growth} in multifarious ways.
Unstable regions are organized into bands separated by finite gaps. Bandwidth
rises with $h$, with the most rapid change around $h=1$. Bands are regularly
spaced at $\omega$\nobreakdash-steps of $\tfrac12$, and are alternately
sign-preserving and sign-reversing (\cref{sec:applied Floquet}). Unstable modes
have $(x,\dot x)$ going clockwise around the origin with winding number
$\tfrac12q$; here $q$ is the band order, which is one for the lowermost band
and one higher for every band above it. The striking similarities between the
two figures are evidence that our toy model captures the essential features of
the parametric bands of eccentric \ac{MRI}, and that what we learn about the
former can provide guidance in understanding the latter.

The left side of the $q$th-order band in \cref{fig:model growth} appears to
pinch off to the point $(h,\omega)=(0,\tfrac12q)$. To see whether this is true,
we determine band edges at $h\ll1$ by solving \cref{eq:model} perturbatively
(\cref{sec:parametric resonance}), as was previously done in the context of
parametric resonance \citep[e.g.,][]{1969mech.book.....L}. The perturbative and
numerical results are in excellent agreement for $q\le5$, confirming that bands
do stretch all the way to the $\omega$\nobreakdash-axis. The width of the
$q$th\nobreakdash-order band, to leading order in $h$, is just
$2^{-3q}q^{2q-1}[(q-1)!]^{-2}h^q$ \citep{1957PGMA....3..132F!}, hence the
figure cannot resolve its extremely thin tip at $h\ll1$ if $q\ge3$. Since bands
extend the instability at $0<h\ll1$ due to parametric resonance to finite $h$,
they and their counterparts in eccentric \ac{MRI} deserve the name
\textquote{parametric bands}.

\Cref{fig:model growth} divides into the small-amplitude regime at $h\le1$ and
the large-amplitude regime at $h>1$. The small-amplitude regime is inherently
stable because $1+h\cos t\ge0$ throughout the period; parametric resonance can
be excited only if $2\omega$ closely matches an integer. In contrast, the
large-amplitude regime is inherently unstable because $1+h\cos t<0$ over part
of the period, which allows exponential growth for a finite amount of time.

\subsection{Implications for eccentric \texorpdfstring{\ac{MRI}}{MRI}}
\label{sec:implications}

The toy model suggests that parametric bands in eccentric \ac{MRI} should be
understood as the result of orbital variation coupling to magnetic oscillation.
At small $e$, weak orbital variation slightly modulates the frequencies of
stable circular modes (\cref{sec:limit}). The physics of eccentric \ac{MRI} in
this regime is the same as parametric resonance: Almost all of the
$(e,f)$\nobreakdash-space at $e\ll1$ and $f^2\ge3$ is stable, with instability
restricted to ranges around discrete values of $f$, ranges that shrink rapidly
as $f$ increases. This tight constraint on $f$ is because parametric resonance
demands a close frequency match. At large $e$, strong orbital variation
overwhelms stable circular modes to the degree that exponential growth is
possible over a part of the orbit near pericenter. This phenomenon can be
viewed as an extension of parametric resonance to large $e$, but it is opposite
to parametric resonance in terms of the $f$\nobreakdash-range that is stable:
Now almost all of $(e,f)$\nobreakdash-space is unstable, whereas stability
requires a close frequency match.

While there is a visual resemblance only between parametric bands in
\cref{fig:MRI growth} and bands in \cref{fig:model growth}, it is easy to infer
that horns are also part of the parametric phenomenon; after all, they derive
from the same \cref{eq:reduced MRI} with merely a different $g$. Parametric
behavior in \ac{MRI} involves gradually increasing degree of destabilization of
stable circular modes as $e$ increases from zero, so its defining feature must
be that each of its unstable regions narrows leftward and collapses to a point
on the $f$\nobreakdash-axis; parametric bands and horns do just that, as we now
show. Band and horn modes in the circular limit obey \cref{eq:dispersion}. Our
toy model suggests that parametric resonance occurs for both bands and horns
around $\omega/n=\tfrac12q$ for some integer $q\ge1$; the
$f$\nobreakdash-coordinates satisfying this condition are marked separately for
band and horn modes with gray triangles in \cref{fig:MRI growth}. Each band and
horn clearly converges toward its respective gray triangle as $e\to0$. It turns
out that $q$ here equals the $q$ denoting band or horn order in
\cref{sec:phase}: The destabilization of a stable circular mode executing
$\tfrac12q$ oscillations per orbit should produce unstable modes with the same
winding number $\tfrac12q$ and thus order $q$.

We can estimate the threshold between the small- and large-amplitude regimes as
follows. If we keep only terms proportional to $\odds{(\lambda v^\eta)}M$ and
$\lambda v^\eta$ in the second row of \cref{eq:MRI oscillator}, we have
\begin{equation}\label{eq:simplified MRI oscillator}
\odd{}M(\lambda v^\eta)+\abbrev J(\lambda v^\eta)\approx0.
\end{equation}
This simplification was used by \citet{1998RvMP...70....1B}, and its validity
is justified by its results. Because $\abbrev J$ is a periodic function, this
equation describes an oscillator with frequency modulated over $M$. The
properties of the oscillator depend on how the average or baseline of $\abbrev
J$ compares with its amplitude. Observe that
\ifapj
  \begin{equation}
  \abbrev J=
    f^2-\frac{3-2e^2}{(1-e^2)^3}-\frac{7e-4e^3}{(1-e^2)^3}\cos\theta
    -\frac{5e^2-2e^4}{(1-e^2)^3}\cos^2\theta-\frac{e^3}{(1-e^2)^3}\cos^3\theta.
  \end{equation}
\else
  \begin{multline}
  \abbrev J=
    f^2-\frac{3-2e^2}{(1-e^2)^3}-\frac{7e-4e^3}{(1-e^2)^3}\cos\theta \\
    -\frac{5e^2-2e^4}{(1-e^2)^3}\cos^2\theta-\frac{e^3}{(1-e^2)^3}\cos^3\theta.
  \end{multline}
\fi
We take the baseline of $\abbrev J$ to be the first two terms on the right-hand
side. Since the coefficients of all powers of $\cos\theta$ are negative, we
approximate the amplitude of $\abbrev J$ to be the negative of their sum. In
the toy model, the transition from the small- to the large-amplitude regime
happens when the amplitude of the frequency modulation equals the baseline
frequency, that is, when $h=1$. In eccentric \ac{MRI}, we expect the same
transition when the amplitude of $\abbrev J$ equals its baseline, that is, when
$(1-e)^3f^2=3-2e$; this is the threshold plotted in \cref{fig:MRI growth}, up
to a factor of unity. The assumption we made to arrive at \cref{eq:simplified
MRI oscillator} is of course \foreign{ad hoc}, but the fact that it reproduces
the threshold means that the terms discarded from \cref{eq:MRI oscillator},
including drag-like terms and mass-like cross-terms, do not enter into the
essence of eccentric \ac{MRI}.

The threshold between the mostly stable and mostly unstable regimes can also be
derived using a more physical argument. \Ac{MRI} grows when orbital shear
stretches out magnetic field perturbations; larger $f$ makes the background
magnetic field stiffer, restricting the region where orbital shear operates to
a smaller fraction of the orbit near pericenter. Circular \ac{MRI} grows if
$f=kv_\su A/n<\smash{\sqrt3}$; by analogy, eccentric \ac{MRI} should grow if
$kv_\su A/(n\Omega)<\smash{\sqrt3}$ at pericenter, that is, if
$(1-e)^3f^2<3+3e$. This is very similar to the criterion just derived.

We close this section with an insight regarding eccentric \ac{MRI} that follows
from a contrast between our toy model and the full picture. Eccentric \ac{MRI}
has a classical band at $0<f^2\lesssim3$, but the toy model does not have a
corresponding band at $0<\omega\lesssim\tfrac12$. This is because the classical
band in eccentric \ac{MRI} is the extension of circular \ac{MRI} to $e>0$
(\cref{sec:growth}), but no such extension is possible for the toy model, which
is always stable at $h=0$. Thus, orbital variation can drive \ac{MRI} whether
or not the time-averaged orbital shear can do so.

\section{Discussion}
\label{sec:discussion}

\subsection{Nonlinear and saturated stages of \texorpdfstring{\acs*{MRI}}{MRI}}
\label{sec:saturation}

We have treated only the linear stage of \ac{MRI} in eccentric disks, but
\ac{MHD} stresses in real disks depend on how \ac{MRI} leads to saturated
\ac{MHD} turbulence, both at what rate and to what final amplitude.

On the one hand, eccentric \ac{MRI} has a $\gamma$ that is typically a sizable
fraction of the maximum $\gamma$ of circular \ac{MRI} (\cref{sec:growth}),
hence the number of orbits needed for \ac{MRI} to go from linear to saturated
in eccentric disks may be only a few times that in circular disks. Saturation
levels may nevertheless be lower due to the slower linear growth. On the other
hand, modes with $f^2\ge3$ are linearly stable in circular \ac{MRI}, so energy
can reach those small scales only through the nonlinear, relatively slow,
process of turbulent cascade from larger scales, whereas modes with
$(1-e)^3f^2\lesssim3$ are linearly unstable in eccentric \ac{MRI}
(\cref{sec:implications}) and grow right from the start. Since saturation
requires a steady state to prevail at all scales, the fact that smaller-scale
modes grow sooner in eccentric disks may help \ac{MRI} saturate faster.
Saturation levels may likewise be higher. Whether slower growth or a wider
range of unstable wavenumbers is more important can be determined only by
nonlinear simulations of the saturation process.

We can only speculate on how the saturated stage of \ac{MRI} differs in
eccentric and circular disks. Self-similar turbulence is characterized by two
wavenumbers: a smaller wavenumber corresponding to the scale at which
turbulence is driven and kinetic energy is injected, and a larger wavenumber
corresponding to the scale at which microscopic dissipation converts kinetic
energy to internal energy. The inertial range refers to the range between these
two wavenumbers; the turbulent power spectrum is a power law in this range. In
the linear stage, eccentric \ac{MRI} is unstable up to wavenumbers
$(1-e)^{-3/2}$ times larger than circular \ac{MRI} (\cref{sec:implications});
in the saturated stage, it is plausible that the driving range reaches
similarly large wavenumbers. However, because small-scale dissipation is
independent of large-scale motion, the inertial range should always cut off at
about the same wavenumber, so we expect the inertial range at $e>0$ to be
narrower. For fixed mean motion and vertically integrated pressure, the rate of
energy injection at $e>0$ may be higher, and the power-law index of the
inertial range may be different. It is also possible that band modes with
dominant magnetic components could interact nonlinearly with horn modes with
dominant velocity components (\cref{sec:cylindrical}), leading to quantitative
changes in \ac{MHD} turbulence, especially when $e\approx1$.

\subsection{Additional physics}

So far we have considered incompressible eccentric \ac{MRI} assuming vertical
wavevectors and ignoring vertical gravity; we interpret its unstable modes
either as stable circular modes destabilized at $0<e\ll1$ by orbital variation
through parametric resonance (\cref{sec:implications}), or as inertial and
sliding modes destabilized at $f>0$ (\cref{sec:limit}). If we allow for
non-vertical wavevectors and vertical gravity, then parametric resonance in
hydrodynamic disks can also destabilize inertial and gravity modes
\citep{2005A&A...432..743P}. We may therefore expect \ac{MHD} disks to
generally host destabilized and magnetically modified inertial, sliding, and
gravity modes.

The height of thin eccentric disks responds to the modulation of vertical
gravity along the orbit, and even mildly eccentric disks can thicken
dramatically from pericenter to apocenter \citep{2014MNRAS.445.2621O}. Vertical
oscillation has the same timescale as orbital variation. The two cooperate in
hydrodynamic disks to destabilize inertial modes through parametric resonance
\citep{2014MNRAS.445.2637B}; the same may happen to the three aforementioned
modes in \ac{MHD} disks.

Vertical oscillation also changes the background density $\rho$, as well as the
vertical wavenumber $k$ of a mode advected with the flow. The resulting
modulation of $v_\su A=B/\rho^{1/2}$ and $f=kv_\su A/n$ means modes may switch
between stable and unstable within an orbit as the disk shuttles between small-
and large-amplitude regimes (\cref{sec:implications}). Moreover, stable
circular modes may parametrically resonate with orbital motion in a different
manner because their $f$ is no longer constant.

Lastly, the background flow of eccentric disks may vary in eccentricity and
orientation as a function of semilatus rectum. Horizontal compression and
expansion of the background flow may alter inertial, sliding, and gravity
modes; it also changes $\rho$ and thus $f$ of stable circular modes.

\subsection{Implication for \texorpdfstring{\acsp*{TDE}}{TDEs}}
\label{sec:TDE}

Our work sheds light on the evolution of the bound debris of \acp{TDE} around
\acp{SMBH}. The debris typically has $e\gtrsim0.99$, so eccentric \ac{MRI} can
grow for values of $f$ that are $\mathrelp\sim(1-e)^{-3/2}\gtrsim1000$ times
greater than in the circular limit (\cref{sec:implications}). With such a broad
$f$\nobreakdash-range linearly unstable, saturation of \ac{MRI}-driven \ac{MHD}
turbulence may take place in only a few orbits (\cref{sec:saturation}), so
angular momentum transport at the rate associated with a saturated state could
begin with relatively little delay.

Our linear formalism tells us how fast \ac{MRI} amplifies \ac{MHD}
perturbations, not the magnitude of \ac{MHD} stresses at saturation. Improving
on the estimates made by \citet{2017MNRAS.467.1426S} of whether angular
momentum transport or energy dissipation is more efficient requires nonlinear
calculations. Nevertheless, we may expect both effects to be weaker near
apocenter if the oscillatory behavior of \ac{MHD} stresses (\cref{sec:stress})
carries over from the linear to the saturated stage, and if
high\nobreakdash-$f$ modes dominate at saturation.

\Ac{MHD} stresses may also give rise to differential apsidal precession in the
saturated stage, as they do in the linear stage (\cref{sec:stress}). Such
precession spreads the range of apsidal orientation of the debris, perhaps
resulting in weak apocentric shocks. In contrast, \ac{GR} bulk apsidal
precession rotates every debris orbit through an angle inversely proportional
to its pericenter distance. For pericenter distances \num{\gtrsim10} times the
gravitational radius, the precession angle is small enough that stream crossing
occurs near apocenter \citep{2015ApJ...804...85S, 2015ApJ...812L..39D}; for
smaller pericenter distances, large swings may lead to closer-in stream
crossing and strong shocks. It is unclear whether shocks accompanying \ac{MHD}
and \ac{GR} precession enhance or diminish the eccentricity of the orbits
closest to the \ac{SMBH} \citep{2017MNRAS.467.1426S, 2017MNRAS.464.2816B}.

If the eccentricity of the inner parts of the debris rises due to either
angular momentum transport or shocks, they will plunge directly across the
innermost stable circular orbit even though they have lost little orbital
energy to radiation. Detailed simulations are required to determine under what
circumstances plunging is the likely scenario.

\section{Conclusions}

We have demonstrated that our intuitions regarding circular \ac{MRI} carry over
to eccentric \ac{MRI}. Orbital shear amplifies the perturbation along those
parts of the orbit where it dominates background magnetic field tension
(\cref{sec:phase}); when it does, it correlates the horizontal components of
velocity and magnetic field perturbations, which leads to radial transport of
angular momentum and energy (\cref{sec:stress}). If we consider growth over the
entire orbit, the perturbation grows if $(1-e)^3f^2\lesssim3$
(\cref{sec:implications}); consequently, \ac{MRI} may be relevant in eccentric
disks, such as the bound debris of \acp{TDE} (\cref{sec:TDE}), up to much
stronger magnetic fields for a given sound speed.

What distinguishes eccentric \ac{MRI} from circular \ac{MRI} is that orbital
conditions vary with time in the former, not in the latter. At small $e$, weak
orbital variation interacts with stable circular modes through parametric
resonance; the whole $(e,f)$\nobreakdash-space is stable except for where
orbital motion resonates with magnetic oscillation. At large $e$, orbital
variation overcomes magnetic oscillation and enables exponential growth; the
whole $(e,f)$\nobreakdash-space is unstable except at resonance
(\cref{sec:implications}).

\ifapj\acknowledgments\else\vskip\bigskipamount\noindent\fi
This research was partially supported by NASA/ATP grant NNX14AB43G, NSF grant
AST-1516299, ERC advanced grant \textquote{TReX}, and ISF I-CORE
\textquote{Origins}. J.H.K. thanks the Kavli Institute for Theoretical Physics
(KITP) for its hospitality during the initiation of this project, and for the
support provided by KITP under NSF grant PHY-1125915.

\begin{appendices}

\section{Eccentric MRI at band and horn edges}
\label{sec:edge}

\Cref{eq:shearing-box MRI} can be integrated analytically when $f=0$. The
non-vanishing elements of the principal fundamental matrix $\mat G(M)$ are
\begin{align}
G_{11}(M) &= \frac{e+\cos\theta}{1+e}, \\
G_{12}(M) &= \frac{2\sin\theta}{1+e}, \\
G_{21}(M) &= -\frac{\sin\theta(1+e\cos\theta)}{2(1+e)}, \\
G_{22}(M) &= \frac{\cos\theta(1+e\cos\theta)}{1+e}, \\
G_{33}(M) &= 1, \\
\nonumber G_{43}(M) &= \frac{3e\sin\theta(1+e\cos\theta)}{2(1-e^2)} \\
&\ifapj\cont\fi\ifarxiv\kern.7em\relax\fi\iflocal\kern-1.4em\relax\fi
  -3\Omega\biggl\{
  \arctan\biggl[\left(\frac{1-e}{1+e}\right)^{1/2}\tan\frac\theta2\biggr]
  +\pi\ceil*{\frac\theta{2\pi}-\frac12}\biggr\}, \\
G_{44}(M) &= \biggl(\frac{1+e\cos\theta}{1+e}\biggr)^2.
\end{align}
We have only three modes because $\mat G(2\pi)$ is non-diagonalizable. The
modes happen to be the first, second, and fourth columns of $\mat G(M)$, and
their Floquet multipliers are all unity, in agreement with $\gamma=0$ along the
$e$\nobreakdash-axis in \cref{fig:MRI growth}. The third column is not a mode
because, instead of all components increasing by the same factor from
pericenter to pericenter as in \cref{eq:mode}, $v^\xi$, $\lambda v^\eta$, and
$w^\xi$ are time-independent while $\lambda w^\eta$ accrues a constant amount
$-3\pi(1+e)^{1/2}(1-e)^{-3/2}$ every orbit due to orbital shear stretching out
radial magnetic field perturbations. The intriguing result here is that, when
$\mat G(2\pi)$ is non-diagonalizable, a suitably initialized perturbation can
grow despite $\gamma=0$, and growth is linear insofar as only pericenter values
are concerned.

This \definition{quasilinear growth} is quite general. Numerical
experimentation reveals that all band and horn edges have non-diagonalizable
$\mat G(2\pi)$ and vanishing $\gamma$; the $f=0$ limit above is simply the
lower edge of the classical band. A perturbation undergoes quasilinear growth
only precisely at an edge; however, because $\mat G(2\pi)$ varies smoothly over
$(e,f)$, the same perturbation grows by a similar magnitude in the neighborhood
of the edge as well. This means a perturbation can grow, at least for a limited
time, faster than what the small near-edge $\gamma$ would indicate.

\section{Parametric resonance in toy model}
\label{sec:parametric resonance}

Suppose the unstable modes of \cref{eq:model} have the form
\begin{equation}
x(t)\eqdef a_0(t)+\sum_{j=1}^\infty a_j(t)\cos\tfrac12jt
  +\sum_{j=1}^\infty b_j(t)\sin\tfrac12jt.
\end{equation}
This ansatz is justified because such modes are either sign-preserving or
sign-reversing (\cref{sec:applied Floquet}). Clearly
\begingroup
\ifarxiv\medmuskip.15\medmuskip\thickmuskip.25\thickmuskip\fi
\iflocal\medmuskip.45\medmuskip\thickmuskip.7\thickmuskip\fi
\begin{align}
\dot x &= \dot a_0
  +\sum_{j=1}^\infty(\dot a_j+\tfrac12jb_j)\cos\tfrac12jt
  +\sum_{j=1}^\infty(\dot b_j-\tfrac12ja_j)\sin\tfrac12jt, \\
\ddot x &= \ddot a_0
  +\sum_{j=1}^\infty(\ddot a_j+j\dot b_j-\tfrac14j^2a_j)\cos\tfrac12jt
  +\sum_{j=1}^\infty(\ddot b_j-j\dot a_j-\tfrac14j^2b_j)\sin\tfrac12jt.
\end{align}\endgroup
Substituting these into \cref{eq:model} furnishes us with
\begin{multline}
(\ddot a_0+\omega^2a_0+\tfrac12h\omega^2a_2) \\
+[\ddot a_1+\dot b_1+(\omega^2-\tfrac14)a_1+\tfrac12h\omega^2(a_1+a_3)]
  \cos\tfrac12t \\
+[\ddot b_1-\dot a_1+(\omega^2-\tfrac14)b_1-\tfrac12h\omega^2(b_1-b_3)]
  \sin\tfrac12t \\
+[\ddot a_2+2\dot b_2+(\omega^2-1)a_2+\tfrac12h\omega^2(2a_0+a_4)]\cos t \\
+[\ddot b_2-2\dot a_2+(\omega^2-1)b_2+\tfrac12h\omega^2b_4]\sin t \\
+\sum_{j=3}^\infty[\ddot a_j+j\dot b_j+(\omega^2-\tfrac14j^2)a_j
  +\tfrac12h\omega^2(a_{j-2}+a_{j+2})]\cos\tfrac12jt \\
+\sum_{j=3}^\infty[\ddot b_j-j\dot a_j+(\omega^2-\tfrac14j^2)b_j
  +\tfrac12h\omega^2(b_{j-2}+b_{j+2})]\sin\tfrac12jt=0.
\end{multline}
All Fourier coefficients must independently vanish. We are interested in
solutions of the form $a_j(t),b_j(t)\propto e^{st}$, hence
\begin{alignat}{2}
(s^2+\omega^2)a_0+\tfrac12h\omega^2a_2 &= 0, \\
(s^2+\omega^2-\tfrac14)a_1+sb_1+\tfrac12h\omega^2(a_1+a_3) &= 0, \\
(s^2+\omega^2-\tfrac14)b_1-sa_1-\tfrac12h\omega^2(b_1-b_3) &= 0, \\
(s^2+\omega^2-1)a_2+2sb_2+\tfrac12h\omega^2(2a_0+a_4) &= 0, \\
(s^2+\omega^2-1)b_2-2sa_2+\tfrac12h\omega^2b_4 &= 0, \\
(s^2+\omega^2-\tfrac14j^2)a_j+jsb_j+\tfrac12h\omega^2(a_{j-2}+a_{j+2}) &= 0,
  &\quad& j\ge3, \\
(s^2+\omega^2-\tfrac14j^2)b_j-jsa_j+\tfrac12h\omega^2(b_{j-2}+b_{j+2}) &= 0,
  && j\ge3.
\end{alignat}
Consider the case when $h\ll1$ and $\omega\eqdef\tfrac12q+\epsilon$, where $q$
is a positive integer and $\abs\epsilon\ll1$. We make the standard assumption
that $s$ and $\epsilon$ are of the same order
\citep[e.g.,][]{1969mech.book.....L}; to first order of $\epsilon$, we have
\begin{alignat}{2}
A_{q,0}a_0+\tfrac18hq^2a_2 &= 0, \\
A_{q,1}a_1+sb_1+\tfrac18hq^2(a_1+a_3) &= 0, \\
A_{q,1}b_1-sa_1-\tfrac18hq^2(b_1-b_3) &= 0, \\
A_{q,2}a_2+2sb_2+\tfrac18hq^2(2a_0+a_4) &= 0, \\
A_{q,2}b_2-2sa_2+\tfrac18hq^2b_4 &= 0, \\
A_{q,j}a_j+jsb_j+\tfrac18hq^2(a_{j-2}+a_{j+2}) &= 0, &\quad& j\ge3, \\
A_{q,j}b_j-jsa_j+\tfrac18hq^2(b_{j-2}+b_{j+2}) &= 0, && j\ge3,
\end{alignat}
where
\begin{equation}
A_{q,j}\eqdef
  \begin{cases} q\epsilon, & j=q, \\ \tfrac14(q^2-j^2), & j\ne q. \end{cases}
\end{equation}
We determine $\epsilon$ at which the oscillator is neutrally stable by setting
$s=0$ in the equations and demanding that they have a non-trivial solution for
the now time-independent $a_j$ and $b_j$. The equation splits into two
independent sets, one involving only $a_j$, the other involving only $b_j$,
hence there are two solutions for $\epsilon$. For $q=1$, the solution is well
known \citep[e.g.,][]{1969mech.book.....L}. For $q\ge2$, the equations are
self-consistent if, to leading order,
\begin{align}
\epsilon &\sim h^2, \\
a_j,b_j &\sim \begin{cases} h^{\abs{j-q}/2}, & (j-q)\bmod2\equiv0, \\
  0, & (j-q)\bmod2\equiv1; \end{cases}
\end{align}
symmetry suggests that we truncate each set of equations at $j=2q$. For all
$q$, we solve for $\epsilon$ up to the lowest order in $h$ such that the two
solutions are distinct; this yields \citep[see also][]{1957PGMA....3..132F!}
\begin{equation}\label{eq:model solution}
\epsilon(h)\approx\sum_{j=1}^{\floor{q/2}}B_{q,j}h^{2j}
  \pm2^{-3q}q^{2q-1}[(q-1)!]^{-2}h^q,
\end{equation}
where $B_{q,j}$ for the first few $q$ are given in \cref{tab:coeffs}.

\begin{table}
\caption{Coefficients in \cref{eq:model solution}.}
\label{tab:coeffs}
\begin{tabular}{ccccc}
\toprule
$q$ & $B_{q,1}$ & $B_{q,2}$ & $B_{q,3}$ & $B_{q,4}$ \\\midrule
2 & $\tfrac{1}{12}$ \\[\smallskipamount]
3 & $\tfrac{27}{256}$ \\[\smallskipamount]
4 & $\tfrac{2}{15}$ & $\tfrac{22}{225}$ \\[\smallskipamount]
5 & $\tfrac{125}{768}$ & $\tfrac{1328125}{8257536}$ \\[\smallskipamount]
6 & $\tfrac{27}{140}$ & $\tfrac{159651}{627200}$ &
  $\tfrac{175093407}{702464000}$ \\[\smallskipamount]
7 & $\tfrac{343}{1536}$ & $\tfrac{9058973}{23592960}$ &
  $\tfrac{1520564265367}{3261490790400}$ \\[\smallskipamount]
8 & $\tfrac{16}{63}$ & $\tfrac{11008}{19845}$ & $\tfrac{57073664}{68762925}$ &
  $\tfrac{263023869952}{238263535125}$ \\[\smallskipamount]
9 & $\tfrac{729}{2560}$ & $\tfrac{779623947}{1009254400}$ &
  $\tfrac{559841590208961}{397888454656000}$ &
  $\tfrac{4562059629450856483359}{2039226076726558720000}$ \\
\bottomrule
\end{tabular}
\end{table}

\end{appendices}

\ifapj\bibliography{mri}\fi
\ifarxiv\printbibliography\fi
\iflocal\printbibliography\fi

\end{document}